\documentstyle[12pt,aaspp4,eqsecnum]{article}
\input epsf

\def \ltaprx {\lower .1ex\hbox{\rlap{\raise .6ex\hbox{\hskip .3ex
	{\ifmmode{\scriptscriptstyle <}\else 
		{$\scriptscriptstyle <$}\fi}}}
	\kern -.4ex{\ifmmode{\scriptscriptstyle \sim}\else 
		{$\scriptscriptstyle\sim$}\fi}}}
\def\gtaprx {\lower .1ex\hbox{\rlap{\raise .6ex\hbox{\hskip .3ex
	{\ifmmode{\scriptscriptstyle >}\else 
		{$\scriptscriptstyle >$}\fi}}}
	\kern -.4ex{\ifmmode{\scriptscriptstyle \sim}\else 
		{$\scriptscriptstyle\sim$}\fi}}}

\def\msun{{\rm ~M_\odot}}
\def\sun{$_{\scriptscriptstyle_\odot}$}
\def\yr{{\rm ~yr}}
\def\g{{\rm ~g}}
\def\cm{{\rm ~cm}}
\def\s{{\rm ~s}}
\def\K{{\rm ~K}}
\def\erg{{\rm ~erg}}
\def\et{{\it et al.~}}

\def\sA{{\cal{A}}}
\def\sD{{\cal{D}}}
\def\sE{{\cal{E}}}

\begin{document}

\title{HYPER-ACCRETING BLACK HOLES AND GAMMA-RAY BURSTS}

\author{Robert Popham\altaffilmark{1},
	S. E. Woosley\altaffilmark{1,2},
	and Chris Fryer\altaffilmark{2}}
\altaffiltext{1}{Max Planck Institut f\"ur Astrophysik, Garching,
Germany, 85740; popham@mpa-garching.mpg.de}
\altaffiltext{2}{UCO/Lick Observatory, University of California, Santa Cruz,
  CA 95064; \\ woosley@ucolick.org, cfryer@ucolick.org}

\lefthead{POPHAM, WOOSLEY AND FRYER}
\righthead{BLACK HOLES AND GAMMA-RAY BURSTS}

\date{\today}

\begin{abstract}

A variety of current models for gamma-ray bursts (GRBs) suggest a
common engine - a black hole of several solar masses accreting matter
from a disk at a rate 0.01 to 10 M\sun \ s$^{-1}$. Using a numerical
model for relativistic disk accretion, we have studied steady-state
accretion at these high rates.  Outside about $10^8 \cm$, the disk is
advection dominated; energy released by dissipation is carried in by
the optically thick gas and the disk does not cool. Interior to this
radius, for accretion rates greater than about $0.01 \msun \s^{-1}$, a
global state of balanced power comes to exist between neutrino losses,
chiefly pair capture on nucleons, and dissipation.  As a result of
these losses, the temperature is reduced, the density raised, and the
disk scale height reduced compared to the advective solution. The
sudden onset of neutrino losses (owing to the high temperature
dependence) and photodisintegration leads to an abrupt thinning of the
disk that may provide favorable geometry for jet production.  The
inner disk remains optically thin to neutrinos for accretion rates up
to about $1 \msun \s^{-1}$. Energy emitted in neutrinos is less, and
in the case of low accretion rates, very much less, than the maximum
efficiency factor for black hole accretion (0.057 for no rotation;
0.42 for extreme Kerr rotation) times the accretion rate, $\dot {\rm
M} c^2$. Neutrino temperatures at the last stable orbit range from 2
MeV (no rotation, slow accretion) to 13 MeV (Kerr geometry, rapid
accretion) and the density from $10^{9}$ to $10^{12} \g \cm^{-3}$. The
efficiency for producing a pair fireball along the rotational axis by
neutrino annihilation is calculated and found to be highly variable
and very sensitive to the accretion rate. For some of the higher
accretion rates studied, it can be several per cent or more; for
accretion rates less than 0.05 M\sun \ s$^{-1}$, it is essentially
zero. The efficiency of the Blandford-Znajek mechanism in extracting
rotational energy from the black hole is also estimated. In light of
these results, the viability of various gamma-ray burst models is
discussed and the sensitivity of the results to disk viscosity, black
hole rotation rate, and black hole mass explored. A diverse range of
GRB energies seems unavoidable and neutrino annihilation in
hyper-accreting black hole systems can explain bursts up to 10$^{52}$
erg. Larger energies may be inferred for beaming systems.

\end{abstract}


\section{INTRODUCTION}

Historically the study of black hole and neutron star accretion has
been motivated by the study of active galactic nuclei and accreting
(stellar mass) x-ray sources.  In both cases, accretion is
Eddington-limited to small rates $\dot M \ltaprx ~10^{-8} (M/\msun)
\yr^{-1}$.  For an interesting range of moderate accretion rates, the
energy released by viscous dissipation can be radiated away
efficiently and consequently the disk is thin.

For very low or very high accretion rates, however, the cooling
efficiency of the disk changes and can drop to the point where most of
the dissipated energy is not radiated, but carried into the hole with
the accreting gas in an ``advection-dominated accretion flow" (ADAF).
For low accretion rates, this occurs because the radiative processes
become very inefficient at low density.  The dissipated energy stays
in the gas, which becomes very hot, and this makes the disk
geometrically thick.  Recently, ADAFs corresponding to {\sl low}
accretion rates and optically thin disks have been studied in some
detail (see Narayan, Mahadevan, \& Quataert 1998 for a review).  They
have been used to model both accreting stellar-mass black holes and
the supermassive black hole sources believed to lie at galactic
centers.

Advection-dominated flows also occur for high accretion rates if the
disk is very optically thick.  Radiation does not diffuse out in a
viscous time scale and so again is carried into the hole (Katz 1977;
Begelman 1978).  The solutions resemble, mathematically, the ADAFs at
low density since in both cases the gas cannot cool.

Eventually, however, at extremely high accretion rates, the nature of
ADAFs can change again due to the onset of efficient cooling by
neutrinos. For {\sl neutron stars}, the accretion rate where neutrino
losses begin to dominate photons in the energy loss budget is about
$10^{-4} \msun \yr^{-1}$ (Chevalier 1993; Fryer, Benz, \& Herant
1996).  For black holes, the limit is much higher because energy can
disappear into the hole. As we shall see, it is only for accretion
rates in excess of about $0.01 \msun$ {\sl per second}, that neutrino
radiation can cool the gas accreting into a black hole on a viscous
time scale. We refer to such accretion as ``neutrino dominated''. The
ADAF solutions one obtains neglecting neutrino emission are very
different.

Besides wanting to explore new accretion physics, we are motivated by
a desire to understand GRBs. Most leading models for GRBs ($\S$2)
have in common an engine based upon a ``hyper-accreting black hole'',
a black hole of $\sim 2 - 10$ M\sun \ accreting mass at a rate
sufficient to consume from 0.1 to several solar masses within, at
most, a GRB time scale (on average 20\,s for the long complex class
of bursts (Fishman \& Meegan 1995)).

For the steady state disks we study, the angular momentum in the
initial system defines the radius where the disk forms and the viscous
time scale at that radius gives the approximate duration of the event
(though longer events occur if the disk is continually fed by an
external source such as the collapse of a star). One conclusion of our
paper will be a range of time scales and accretion rates (viscosity
dependent) that should characterize various GRB models. We also
calculate the variation of temperature, density, disk scale height,
radial drift velocity and luminosity with radius in the disk. From
these we are able to infer, albeit approximately, the efficiency for
jet production based on neutrino annihilation (see also Jaroszy\'nski
1996) as well as the energy density in the disk and the number of
orbits experienced by an accreting blob. The latter has implications
for whether the large magnetic fields that have been invoked for the
electromagnetic acceleration of jets can accumulate (Blandford \&
Znajek 1977; Narayan \et 1992; Hartmann \& Woosley 1995; M\'esz\'aros
\& Rees 1997; Paczy\'nski 1998).

Our numerical model is discussed in $\S$3.  We solve the slim disk
equations in the Kerr metric for steady accretion.  This differs from
some previous work which treated the disk as a hydrostatic torus, and
approximated the effects of accretion by a sequence of tori of
decreasing mass (Jaroszy\'nski 1996; Daigne \& Mochkovitch 1997).  The
solutions are characterized by four physical parameters: the black
hole mass, accretion rate, black hole spin parameter $a$, and disk
viscosity parameter $\alpha$. We include, in addition to the emission
from pair annihilation, neutrino losses from the electron-positron
pair capture on nucleons (which we find dominates the cooling), and
photodisintegration. The equation of state, while simple, contains the
effects of radiation, relativistic pairs, and degeneracy.  Our
semi-analytic solutions are thus sufficiently realistic to make
qualitative predictions regarding the viability of various models for
GRBs and to guide the complex multi-dimensional hydrodynamical studies
that will follow (e.g., Eberl, Ruffert, \& Janka 1998; MacFadyen \&
Woosley 1998; Fryer \& Woosley 1998; Fryer et al. 1998).

We discuss the results of our calculations in \S 4 and describe a
typical solution.  We examine solutions for a variety of accretion
rates, disk viscosity parameters, and black hole spin rates.  For each
case, we derive neutrino luminosities, which turn out to depend
strongly on all of these parameters.  Approximations to the accretion
rate are given in \S 5 for various evolutionary scenarios.

In Section 6 we calculate the efficiencies of gamma-ray production by
neutrino annihilation for the various models.  We also estimate the
efficiencies for energy extraction by the Blandford-Znajek process.
The final section summarizes our conclusions and discusses the
implications of our hyper-accreting black hole disk solutions for GRB
models.

\section {EVOLUTIONARY PATHS}

There are many ways in which a black hole may come to experience rapid
transient accretion. We expect that each occurs to an appreciable
extent in nature; indeed given current uncertainties in such
quantities as the neutron star kick velocity, common envelope
evolution and the sizes of stripped helium cores during the late
stages of evolution, it is uncertain just which predominates (Fryer,
Woosley, \& Hartmann 1998). It is likely, though, that GRBs are a
diverse set of phenomena, all having at their heart a hyper-accreting
black hole, but differing in accretion rate, accretion mass, rotation
rate of the black hole, and surroundings (Table 1).

\subsection{Merging Neutron Stars}

The oldest of the binary compact object models for cosmological GRBs,
merging neutron stars, were first mentioned as possible sources by
Paczy\'nski (1986) and Eichler \et (1989).  Progenitor double neutron
star systems have long been known to exist (Hulse \& Taylor 1975) and
several evolutionary paths for their formation have been proposed
(e.g. Srinivasan \& van den Heuvel 1982; Brown 1995).  For a time it
was believed that the burst might originate during the merger, but
this proved too inefficient (Janka \& Ruffert, 1996). However,
depending upon the equation of state, a result of the merger may be a
black hole of about 2.5 solar masses surrounded by a disk of
approximately $0.1 - 0.2 \msun$ (Ruffert \& Janka 1998).

That accretion from this disk might give rise to GRBs was suggested by
Narayan \et (1992). The most detailed calculations to date are by
Davies \et (1994), Ruffert, Janka, \& Sch\"afer (1995), Janka \&
Ruffert (1996), and Ruffert \& Janka (1998). In particular Ruffert \&
Janka (1998) estimate the mass of the disk to be 0.1 - 0.2 M\sun, the
accretion rate about 1 M\sun \ s$^{-1}$, the neutrino luminosity about
10$^{53}$ erg s$^{-1}$, the energy deposition along the polar axes by
neutrino annihilation about $5 \times 10^{50}$ erg s$^{-1}$, and thus
the total energy available to make the burst about $5 \times 10^{49}$
erg. Simple equations derived in \S 5 and efficiencies calculated in
$\S$6 agree well with these numbers.

\subsection {Neutron Stars Merging With Black Holes}

Paczy\'nski (1991) and Narayan \et (1992) also
suggested that the merger of a neutron star with a pre-existing black
hole of several solar masses might produce GRBs.  In the dominant 
formation scenario of binaries consisting of a black hole and 
neutron star, the black hole is formed via hypercritical 
accretion during a common envelope phase (Bethe \& Brown 1998).  
The resultant black hole has a very low mass ($\sim 3 M_\odot$), 
very similar to the double neutron star systems.  Here the situation
is somewhat improved because of the larger mass of the disk that
forms, about $0.5 \msun$ (Eberl 1998; Eberl, Ruffert, \& Janka 1998;
Klu\'{z}niak \& Lee 1998) and the fact that one already has a black
hole to start with. In most ways, though, this is just a more
energetic analogue of the merging neutron star aftermath.  Accretion
rates and neutrino luminosities are higher because of the larger disk
mass. The neutrino annihilation efficiency is consequently increased
to a few per cent. Eberl \et (1998) find disk masses of 0.5 M\sun,
accretion rates highly variable between 1 and 10 M\sun \ s$^{-1}$,
neutrino luminosities from 10$^{53}$ to 10$^{54}$ erg s$^{-1}$, and
efficiencies for neutrino annihilation of a few per cent. The total
energy available for the burst is about 10$^{51}$ erg. Depending upon
beaming, this may still be inadequate to explain GRB 971214 (Ramaprakash
\et 1998) without invoking energy sources other than neutrino
annihilation, but it may account well for the short hard bursts in the
BATSE sample (Fishman \& Meegan 1995). A concern both for this model
and the merging neutron star model is the distance the system may
travel before merging, perhaps too far to be associated with star
formation.  However, because the trigger threshold on Beppo-Sax
exceeds a few seconds, we have no information on the optical
counterparts to short GRBs.

\subsection {Collapsars}

The mechanism whereby the collapsing iron core of a massive star gives
rise to an outgoing shock and makes a supernova remains controversial
(e.g., Mezzacappa \et 1997; Fryer 1998).  Surely many massive stars,
in their deaths, produce supernovae by forming neutron stars in
spherically symmetric explosions, but perhaps not all do. Especially
as the mass of the star becomes more than $\sim 30 \msun$ on the main
sequence (pre-supernova helium core masses bigger than 12 M\sun), the
accretion rate of the mantle onto the proto-neutron star becomes so
great, $\sim 1 \msun \s^{-1}$, and the mass of the iron core so large
($\gtaprx ~2.0 \msun$), that a black hole may form before a neutrino
powered shock is successfully launched.  Woosley (1993, 1996)
suggested that the outcome of such collapses would be gamma-ray
bursts.  This model is currently being explored in two-dimensional
calculations by MacFadyen \& Woosley (1998) and a very similar model,
named the ``hypernova", has been discussed by Paczy\'nski (1997). An
explosion of this sort may have been associated with Supernova 1998bw
(Woosley, Eastman, \& Schmidt 1998).

Early on, material accretes rapidly though the poles and the angular
momentum is so low that accretion through the equator is also almost
unimpeded.  The hole rapidly grows to over 3 M\sun. But as the
angular momentum of the accreting material rises to 
0.5 - 1 $\times 10^{17}$ cm$^2$ s$^{-1}$, a portion of the
remainder accretes through a disk. Two-dimensional calculations by
MacFadyen \& Woosley (1998) show that, for a 14 M\sun \ helium star
after about 10 seconds, the polar regions are evacuated while disk
accretion continues at over 0.1 $\msun \s^{-1}$ for about 10 s (with a
subsequent long, slow decline). During most of this accretion the Kerr
parameter of the black hole is over 0.9.

\subsection {White Dwarfs Plus Black Holes}

White dwarfs may also merge with black holes at an appreciable rate
(Fryer et al. 1998).  Except for the smaller secondaries, formation
scenarios for binaries consisting of a white dwarf and black hole
parallel those of neutron stars and black holes.  Fryer, Woosley, \&
Hartmann (1998) find that the white dwarfs formed in these binaries
will preferentially have large masses ($\sim 1 M_\odot$).  These
massive white dwarfs are tidally disrupted at a few times $10^9 \cm$
and form an accretion disk around the black hole.  The accretion into
the hole occurs by way of a disk at a rate of about $0.01 - 0.07 \msun
\s^{-1}$ and lasts about $10 - 70 \s$.  The lower accretion rate and
longer timescale corresponds to low-mass white dwarfs ($M_{\rm WD}=0.7
M_\odot$), whereas the high mass white dwarfs which dominate these
binaries have higher accretion rates, but lower accretion timescales.
Again the black hole is spun up by the accretion (if it were not
already rotating rapidly) to spin parameters of $a \sim 0.5$.

\subsection {Black Holes and Common Envelope Evolution}

In many of the formation scenarios for compact binaries, a common
envelope phase is required to create the short periods.  The more
massive component first makes a black hole or neutron star and is
later enveloped as the less massive star becomes a giant.  A fraction
of these systems will not eject the hydrogen common envelope; instead,
the rapidly accreting compact primary will merge with the secondary's
helium core, accreting quickly enough to become a black hole if it is
not already.  The accretion of the helium core onto the black hole has
been proposed as a GRB progenitor (Fryer \& Woosley 1998).  The helium
core is disrupted by the black hole into an accretion disk with a
radius equal to a fraction of the initial core, $\sim 10^9 - 10^{10}
\cm$.  The accretion rate, especially along the poles, can be quite
large, initially perhaps $1 \msun \s^{-1}$, but again the disk
accretion rate is expected to be smaller (and last longer).

Detailed calculations do not exist, but the angular momentum is larger
than in the collapsar model, comparable in fact to the white dwarf--
black hole merger. The disk will thus form at a large radius,
comparable to that of the helium core. A crude estimate is 1 - 10
$\times 10^9$ cm. At this distance the viscous time scale will range
from a fraction of a minute to 10 minutes (\S 5; Table 1) and the
accretion rate spans a comparably large range. Again the black hole
may accrete an amount comparable to its mass and be spun up to high
Kerr parameters. 

\section{COMPUTATIONAL PROCEDURE}

\subsection {The Physical Model}

The evolutionary scenarios described above all produce rapid disk
accretion ($0.01-10 \msun \s^{-1}$) onto a black hole of a few solar
masses.  For present purposes, we are interested in the gross
properties of these disks rather than their detailed time-dependent
behavior; so, for simplicity, we use a steady-state disk model to
study this transient event.  This should be a reasonable
approximation, since the viscous time scales in the inner disk, which
should have the highest temperature and neutrino flux, are much
shorter than those in the outer disk, on which $\dot M$ should vary.

As discussed in $\S 1$, we expect that the disk will be unable to cool
efficiently via photons, and so the energy dissipated in the disk will
be advected inward.  When the temperature and density are sufficiently
high, neutrino cooling will become efficient.  Thus our model for the
disk includes advection, but also allows for cooling which varies
substantially with disk radius, unlike most previous models of ADAFs.
Also, because the inner regions of the disk and the spin of the black
hole may be important, we solve for the structure of the disk in the
Kerr metric.

Our disk model is based on the advection-dominated disk model of
Gammie \& Popham (1998, hereafter GP) and Popham \& Gammie (1998,
hereafter PG).  This model solves the ``slim disk'' equations in Kerr
geometry.  The slim disk equations are a more sophisticated version of
the standard thin disk equations which include radial pressure
gradients and radial energy transfer.  For a detailed description of
these equations and the form they take in the Kerr metric, we refer
the reader to GP; however, we review them here for the reader's
convenience. 

The units are such that $G = M = c = 1$, where $M$ is the mass
of the black hole.  Later, when we present our results, we plot the
variables scaled to cgs units.

The continuity equation is:
\begin{equation}
4\pi r \rho H V \left(\sD\over{1 - V^2}\right)^{1/2}
	= -\dot{M}.
\end{equation}
Here $r$ is the Boyer-Lindquist radius, $\rho$ is the rest mass
density, $H$ is the disk thickness, $V$ is
the radial velocity measured in a co-rotating frame, $\sD \equiv 1 -
2/r + a^2/r^2$ is a relativistic correction factor, and $\dot M$ is
the rest mass accretion rate.

The gas energy equation is
\begin{equation}
\rho V \left(\sD\over{1 - V^2}\right)^{1/2}
\left({d u\over{d r}} - {p\over{\rho^2}}{d\rho\over{d r}}\right) = \Phi
- \dot q_{tot} \equiv f \Phi
\end{equation}
Here $u$ is the internal energy, $p$ is the pressure, $\Phi$ is the
dissipation function, and $\dot q_{tot}$ is the total cooling rate, as
described below.  The parameter $f$ measures the importance of
cooling; if $f=1$, all the dissipated energy is advected with the
flow, while if $f=0$, cooling and dissipation occur at the same rate.
Most ADAF models have taken $f$ to be constant with radius (usually
$f=1$), but here we compute $f$ directly from the local dissipation
and cooling rates.  Note that if the cooling rate exceeds the
dissipation rate, one can have $f < 0$.

The radial momentum equation is
\begin{equation}
{V\over{1 - V^2}} {d V\over{d r}} = f_r - {1\over{\rho\eta}}
	{d p\over{d r}},
\end{equation}
where
\begin{equation}
f_r \equiv -{1\over{r^2}}{\sA \gamma_\phi^2\over{\sD}}
	(1 - {\Omega\over{\Omega}}_+) (1 - {\Omega\over{\Omega}}_-).
\end{equation}
The $f_r$ term combines the effects of gravity and rotation,
where $\sA \equiv 1 + a^2/r^2 + 2 a^2/r^3$ and $\gamma_\phi^2 = 1 + l^2
(1-V^2) / (r^2 \sA)$, $\Omega = u^\phi/u^t$ is the angular velocity,
and $\Omega_\pm = \pm (r^{3/2} \pm a)^{-1}$.  The radial acceleration,
on the left-hand side, is given by the difference between $f_r$ and the
pressure gradient force, where $\eta$ is the relativistic enthalpy
$\eta \equiv (\rho + p + u)/\rho$.

The angular momentum conservation equation is
\begin{equation}
\dot{M} l\eta - 4\pi H r {t^r}_\phi = \dot{M} j.
\end{equation}
Here $l$ is the specific angular momentum of the accreting gas, $j =
const.$ is the angular momentum accretion rate per unit rest mass
accreted.  The remaining term gives the viscous angular momentum
transport rate, where $t^r_{\phi}$ is the viscous stress.  The
expression for $t^r_{\phi}$ is rather lengthy, the calculation of
it even more so, and the reader is referred to \S 4 of GP for a full
discussion.

The equation of vertical mechanical equilibrium is
\begin{equation}
H^2 = {p\over{\rho\eta \nu_z^2}},
\end{equation}
where $\nu_z$ is an effective vertical frequency.  We adopt the
expression derived by Abramowicz, Lanza, \& Percival (1997) for
$\nu_z$:
\begin{equation}
\nu_z^2 = {l^2 - a^2 (\sE^2 - 1)\over{r^4}},
\end{equation}
where $\sE = -u_t$ is the ``energy at infinity'', which is conserved 
along geodesics. 

One of the radial energy transfer terms included is the advection of
entropy.  This term can become important in situations where the
cooling of the disk is very inefficient, so that the viscous time scale
is shorter than the cooling time scale.  This can occur when the
optical thickness of the disk, as measured from the midplane to the
surface, is either very low or very high.  In models of
advection-dominated disks, a large fraction $f_{adv}$ of the energy
dissipated in the disk is stored as entropy and advected inward with
the accreting gas.  GP and PG assumed that $f_{adv}$ was constant with
radius, and usually took $f_{adv} = 1$.

For the extremely high accretion rates that we are considering, the
optical thickness of the disk is very large.  Thus the effective
temperature is much lower than the midplane temperature, and the
cooling time is long, so it appears likely that the disk is
advection-dominated.  Highly optically thick disks where advection is
important have been modeled for the case of FU Orionis objects (Popham
\et 1996), but even there the accretion rates ($\sim 10^{-4} \msun
\yr^{-1}$) are many orders of magnitude smaller than those considered
in this paper.  If radiation were the only source of cooling
available, the disk would be advection-dominated; however, eventually,
the disk reaches a temperature where neutrino losses become
significant.

\subsection{Neutrino and Photodisintegration Losses}

Cooling by neutrino emission is important in regions of the disk where the
temperature is sufficiently high.  Because of the steep temperature and
density dependence of the neutrino emissivity, we expect the onset of
neutrino cooling to occur fairly abruptly at some transition radius in the
disk.  Outside this radius, cooling is inefficient and the disk is
advection-dominated; inside, neutrinos cool the disk efficiently.  Because of
this, it is no longer feasible to use an $f_{adv}$ which is constant with
$R$.

We include two types of neutrino losses.  The first is neutrino
emission due to pair annihilation.  These are computed from the
results of Itoh \et\ (1989,1990).  A rough approximation to these
results is given by the simple expression
\begin{equation}
\dot q_{\nu, \bar \nu} \simeq 5 \times 10^{33} T_{11}^9 
\erg \cm^{-3} \s^{-1},
\end{equation}
which can be used to compare pair cooling to the other cooling terms.
The second type of neutrino cooling we include is due to {\sl capture} of
pairs on nuclei.  This is computed according to
\begin{equation}
\dot q_{eN} = 9.0 \times 10^{33} \rho_{10} T_{11}^6 X_{nuc} \erg
\cm^{-3} \s^{-1},
\end{equation} 
where $\rho_{10} = \rho/10^{10} \g \cm^{-3}$, $T_{11} = T/10^{11} \K$,
and $X_{nuc}$ is the mass fraction of nucleons (Qian \& Woosley 1996).
To avoid confusing this pair capture loss rate with that from pair
annihilation, we refer to the former as the ``URCA'' cooling.
Note that URCA cooling will dominate when $1.8 \rho_{10} >
T_{11}^3$.

$X_{nuc}$ is zero in the outer disk, but photodisintegration breaks
down alpha particles into neutrons and protons once $T$ reaches about
$10^{10}$ K.  $X_{nuc}$ is given by
\begin{equation}
X_{nuc} = 30.97 \rho_{10}^{-3/4} T_{10}^{9/8} \exp(-6.096/T_{10})
\end{equation}
where this expression gives $X_{nuc} < 1$, and $X_{nuc} = 1$
elsewhere.  The photodisintegration process cools the gas according to
\begin{equation}
\dot q_{phot} = 10^{29} \rho_{10} V {dX \over dr} \erg \cm^{-3}
\s^{-1}
\end{equation}

For accretion rates greater than $\sim 1 \msun \s^{-1}$, the disk 
begins to be optically thick to its own neutrino emission.  We limit 
the actual neutrino emission to the blackbody limit (Mayle 1985):
\begin{equation}
L_{\nu_{\rm x}}={7 \over 8} \sigma_{\rm B} T^4
\end{equation}
where $\sigma_{\rm B}$ is the Stephan-Boltzmann constant.  
This limit corresponds to the Fermi-Dirac blackbody luminosity
emitted from the neutrinosphere and is the same for each 
neutrino flavor ($\nu_x=\nu_e,\nu_{\mu},\nu_{\tau}$).  
This estimate gives rough agreement with the results of 
Eberl \et (1998) and allows us to extend our results to 
accretion rates as high as $10 \msun \s^{-1}$.

\subsection{Equation of State}

GP and PG assumed a simple gas pressure equation of state.  In the
optically thin disks they were considering, radiation pressure is
unimportant, and the densities are low enough that degeneracy pressure
was negligible.

In the present problem, gas, radiation, and degeneracy pressure are
all significant.  Accordingly we write the pressure as
\begin{equation}
P = \rho R T {(1 + 3 X_{nuc}) \over 4} + {11 \over 12} a T^4 + K
\left({\rho \over \mu_e}\right)^{4/3},
\end{equation} where $R$ is the gas
constant, $K = (2 \pi h c / 3)(3 / 8 \pi m_n)^{4/3} = 1.24 \times
10^{15}$, where $m_n$ is the nucleon mass, and $\mu_e$ is the mass per
electron, which we take to be 2.  The three terms represent gas
pressure from nucleons, pressure from radiation and relativistic
electrons and positrons, and relativistic degeneracy pressure from
electrons, respectively.  The corresponding expression for the
internal energy is
\begin{equation}
u = {3 \over 2} R T {(1 + 3 X_{nuc}) \over 4} + {11 \over 4} {a
T^4 \over \rho} + 3 K \left({\rho \over \mu_e}\right)^{1/3}.
\end{equation}
We have assumed that the gas is pure helium
before photodisintegration, hence the factor $(1 + 3 X_{nuc})/4$ to
account for the change in gas pressure.  We have compared the results
of this expression to those produced by a full EOS routine by
Blinnikov, Dunina-Barkovskaya, \& Nadyozhin (1996), 
and we find that the total pressure
agrees to within 10\% for $10^7 \leq \rho \leq 10^{11}$, $10^9 \leq T
\leq 10^{10.9}$.

\subsection{Solution Method}

We set the outer edge of the disk at $10^4$ Schwarzschild radii, or $r
= 2 \times 10^4 \, G M/c^2$, and the inner edge just outside the event
horizon at $r = (1 + \sqrt{1 - a^2}) \, G M/c^2$, where $\sD = 0$.  

At the outer edge we impose two boundary conditions: $\Omega$ and $c_s$
must equal their values in the self-similar advection-dominated
solution of Narayan \& Yi (1994).  
Two other conditions on the flow are provided by the requirement that
the flow pass smoothly through two critical points.  The first is the
sonic point $r_s$, where $|V| \simeq c_s$.  The second is the
``viscous point'' $r_v$ associated with the finite propagation speed
of viscous effects, where $|V| \simeq c_\nu$.  Associated with each
critical point are two conditions that must be satisfied for a smooth
flow, as well as one degree of freedom, the location of the critical
point itself.

The final boundary condition normalizes the density (the density
appears in the basic equations only in the form $d\ln\rho/d r$).  We
now have all the boundary conditions required to solve the four
first-order ordinary differential equations for $V,l,\rho,$ and $T$,
and to find the eigenvalue $j$.

The equations are solved using a relaxation method, as described by GP
and PG.

\section {RESULTS}

Our solutions are characterized by four parameters: the black hole
spin $a$, the viscosity parameter $\alpha$, the mass accretion rate
$\dot M$, and the black hole mass $M$.  The values of these are
determined largely by the evolutionary path which provides the mass to
the disk.  The black hole mass and spin depend in part on the mass and
angular momentum of the progenitor star which originally collapsed to
produce the black hole, but both $M$ and $a$ can be increased
substantially by the rapid accretion we are modeling here.  The mass
accretion rate will depend largely on the evolutionary path which
leads to the rapid accretion.  The viscosity parameter $\alpha$ is
poorly known, particularly under the kinds of temperatures and
densities which are reached in these disks.  Since the accretion rate
is determined by the mass of the disk and the viscous time scale on
which it accretes, there should be a connection between $\dot M$ and
$\alpha$, as discussed in \S 5.

\subsection{A Typical Solution}

In order to describe the general features of our solutions, we begin
by focusing on a ``standard model'' with $a=0$, $\alpha = 0.1$, $\dot
M = 0.1 \msun \s^{-1}$, and $M = 3 \msun$.  This model is shown in
Figures 1 and 2, and repeated in some other figures for comparison
purposes.  In Figure 1, we compare this standard model to two other
solutions.  The first is a pure advection-dominated solution
calculated using the same code, but with all cooling terms turned off,
and including only gas pressure.  This solution is identical to the
one shown by GP and PG, scaled to physical units using the appropriate
values of $M$ and $\dot M$.  The second is an analytic solution for a
thin neutrino-cooled disk described in \S 5, which assumes Keplerian
rotation, and that pair capture dominates the cooling and gas pressure
dominates the pressure.  Note that this analytic solution assumes a
Newtonian gravitational potential.  These solutions represent the two
extremes of no cooling and highly efficient cooling which balances the
viscous dissipation at each radius.  Thus, we expect that the
characteristics of our neutrino-cooled disk will fall between those of
these two solutions.

At large radii, the accretion flow is simply an advection-dominated
flow.  Figure 1 shows that the disk is quite similar to the pure
advection-dominated flow at $\log r = 9$.  Densities and temperatures
are too small for neutrino cooling to be significant, while optical
depths are too large for significant photon cooling.  The approximate
surface density of the disk is $\Sigma = \rho H \sim 10^{14} \g
\cm^{-2}$ at $r = 10^9 \cm$, so the disk should be extremely optically
thick to photons, but optically thin to neutrinos.  Figure 2 shows
that $f=1$ at large radii, and equivalently that the cooling time scale
is much longer than the accretion time scale, so the dissipated energy
is advected inward before it can be radiated away.  The disk is thick,
with $H \sim R$, and has substantially sub-Keplerian rotation,
reflecting the importance of pressure gradients.

The first significant cooling occurs in the region from $\log r = 7.5 -
8.2$, where the nuclei photodisintegrate into nucleons.  The nuclei
absorb $10^{19}$ erg g$^{-1}$; this energy loss is larger than the energy
added by viscous dissipation, and so the entropy of the gas decreases
in this zone.  This is reflected in the negative value of $f$ in this
region (Fig. 2).  The disk becomes thinner, with $H \approx 10^7 \cm$
at $r = 10^{7.5} \cm$, and the rotation is much closer to Keplerian.
Fig. 1 shows that the disk becomes more like the thin neutrino-cooled
disk in this zone.  Fig. 2 shows that the radiation pressure, which
was the largest pressure term at $\log r = 8.2$, drops in importance as
the disk becomes denser and more nucleons are produced, and gas
pressure, which had dominated in the outer disk, is again the largest
term at $\log r = 7.5$.

At $\log r \approx 7.5$, all of the nuclei have photodisintegrated, and the
cooling due to photodisintegration shuts off rather abruptly, as shown
by the sudden increase in $f$ and in the cooling time (Fig. 2).
Neutrino cooling, predominantly by pair capture, begins to become
important here, but at $\log r = 7.5$ it is still well below the
dissipation rate.  As the gas continues to fall inward, the increasing
temperature and density produce a rapid increase in the neutrino
cooling rate, which gradually approaches the dissipation rate.  This
can be seen in the slow decrease in $f$ from one (advection-dominated)
toward zero inside $\log r = 7.5$.  Radiation pressure again becomes
the dominant term in the inner disk.  Note that in the innermost
portion of the disk, the dissipation rate briefly becomes negative
where the stress and shear rate change sign due
to the use of a causal viscosity prescription (see GP for more
details); this produces the feature in $f$ at $\log r = 6 - 6.4$.
Near the horizon, $V$ is approaching $c$ and $\Omega$ drops toward
$\omega \equiv 2 a / (r^3 + a^2 r + 2 a^2) = 0$.

The neutrino luminosity of this solution is $L_\nu = 3.35 \times
10^{51} \erg \s^{-1}$, of which nearly 90\% is from pair capture (see
Table 2 for luminosities).  This is less than 2\% of $\dot M c^2$, a
somewhat lower efficiency than the standard thin disk figure of 5.7\%,
due largely to the fact that much of the dissipated energy goes into
the entropy of the gas and is advected into the black hole.  Note that
$2 \times 10^{51} \erg \s^{-1}$ goes into photodisintegration of the
nuclei.  The neutrino emission from the disk comes predominantly from
between $\log r \sim 6 - 7$.  The disk is thin to neutrinos; $\Sigma
\simeq 10^{16} \g \cm^{-2}$ in the inner disk, while the opacity to
neutrinos is about $10^{-18} \cm^2 \g^{-1}$ giving an optical 
depth to neutrinos of $\sim 0.01$.

\subsection{Solutions for Other $\dot M$, $\alpha$, $a$, and $M$}

Figure 3 shows solutions for $\dot M = 0.01, 0.1, 1, 10 \msun
\s^{-1}$, all with $a=0$, $\alpha = 0.1$, and $M = 3 \msun$.  The
changes in $\dot M$ produce substantial changes in $\rho$ and in the
neutrino luminosity, but the disk height $H$ and radial velocity $V$
show only small changes, and the temperature increases slowly with
increasing $\dot M$.  The density and neutrino luminosity both show a
strong dependence on $\dot M$; this is because the increased density
increases the neutrino luminosity, which makes the disk thinner and
denser.  The $0.01 \msun \s^{-1}$ solution has a very low neutrino
luminosity $L_\nu \simeq 1.5 \times 10^{49} \erg \s^{-1}$, while the
$1 \msun \s^{-1}$ solution has $L_\nu \sim 8.5 \times 10^{52} \erg
\s^{-1}$, a factor of more than 5000 larger.  At low accretion rates,
the density and temperature are simply too low to permit effective
cooling by neutrino emission, and nearly all the dissipated energy,
apart from that lost to photodisintegration, is advected into the
hole.  At higher accretion rates, the high temperatures and densities
allow the dissipated energy to be radiated away efficiently.  In fact,
both the 1 and 10 $\msun \s^{-1}$ solutions have neutrino luminosities
in excess of the dissipation rate, because the neutrino emission
radiates away some of the gas entropy in addition to the dissipated
energy.  This can be seen in the variations in $f$ in Fig. 3; the
$0.01 \msun \s^{-1}$ solution has $f \simeq 1$ inside of the
photodisintegration zone, while for $0.1 \msun \s^{-1}$ $f$ drops
gradually, reflecting the increasing fraction of the dissipated energy
which is radiated away by neutrinos.  The higher $\dot M$ solutions
both quickly reach $f \ltaprx ~0$ in this zone, demonstrating that
neutrino emission is radiating away more energy than the disk is
dissipating.

Figures 4 and 5 illustrate the effects of changes in $\alpha$.  Each
shows solutions with $\alpha = 0.1$, 0.03 and 0.01; in Fig. 4 $\dot M = 0.1
\msun/ \s$, while Fig. 5 $\dot M = 0.01 \msun \s^{-1}$.  All have
$a=0$, $M = 3 \msun$.  In both cases, decreasing $\alpha$ produces a
solution with lower radial velocity and higher density.  This results
in more efficient neutrino cooling, with higher neutrino luminosities
for the same $\dot M$.  In the $0.01 \msun \s^{-1}$ solutions
(Fig. 5), the change in $\alpha$ produces a dramatic change in
$L_\nu$, since the $\alpha = 0.1$ solution is very inefficient.  The
more efficient cooling of the $\alpha = 0.01$ solutions occurs in the
inner disk, so the temperature and disk height fall below the $\alpha
= 0.1$ solutions there.

One interesting feature of the $\alpha = 0.01$ solutions is that they
show maxima in the density, temperature, and pressure at around $\log
r = 6.6$.  These are similar to the maxima seen by PG at a similar
position in the disk for a solution with $\alpha = 0.001$ and a
solution with $f=0.01$.  Solutions with small values for $\alpha$
and/or $f$ have relatively small radial velocities outside of this
radius, and build up high densities and pressures there.  Inside of
the pressure maxima, the gas falls in toward the hole, with large
radial acceleration.  This produces the drop in temperature and
density, and therefore in neutrino cooling, seen in Figs. 4 and 5.
This effect can also be seen, to a lesser extent, in the $\alpha=0.03$
solutions and in the high-$\dot M$ solutions in Fig. 3.

Figure 6 shows the effects of increasing the black hole spin $a$ for
solutions with $a = 0, 0.5, 0.95$.  Changes in $a$ have little effect
at large radius.  Near the hole, the high-$a$ solutions have higher
densities and stronger neutrino cooling.  The horizon also moves to
smaller radius as $a$ increases.  In the $a=0.5$ solution, the
neutrino luminosity due to annihilation rises dramatically from the
$a=0$ solution.  This results from a small increase in temperature,
since the annihilation luminosity varies approximately as $T^9$.  It
appears that here cooling due to annihilation is acting as a
thermostat which prevents $T$ from rising any further.  In the
$a=0.95$ solution, most of the increase in neutrino luminosity comes
from pair capture, due to the dramatic increase in density in the
inner part of the disk.

Figure 7 shows solutions with $a=0.5$ for four accretion rates: $\dot
M = 0.01$, 0.1, 1.0, and $10 \msun \s^{-1}$.  This can be compared to
Figure 3, where $a=0$ solutions were shown for the same accretion
rates.  The $a=0.5$ solutions have higher temperatures and densities,
and produce higher neutrino luminosities.  These solutions might be
appropriate to the later stages of a burst event, when a substantial
amount of mass and angular momentum has been added to the black hole.

Figure 8 shows how $a$ increases as a black hole accretes mass and
angular momentum from a thin disk; the rates at which the black hole
mass and angular momentum increase are given by Bardeen (1970).  If
the black hole is spun up by accretion from an ADAF, it will spin up
more slowly, and will reach an equilibrium value of $a$ where it
neither spins up nor spins down; see PG for details.  Spinup due to
accretion from a neutrino-cooled disk will be intermediate between the
thin disk and the ADAF cases.

We have also calculated solutions with larger values of the black hole
mass.  These include a $6 \msun$ black hole with $a=0.95$ and a $10
\msun$ hole with $\dot M= 0.01 \msun \s^{-1}$ and $\alpha=0.01$.
These solutions are listed in Table 2, where it can be seen that they
have much lower neutrino luminosities than $3 \msun$ black holes with
the same values of $\dot M$, $a$, and $\alpha$.  This is due to the
larger size of the more massive holes; this makes the density of the
accreting gas lower, and thus reduces the pair capture luminosity.  

\section{APPROXIMATIONS TO THE ACCRETION RATE}

For those models where the accretion rate is governed by the viscous
time scale of the disk, it is possible to make a simple estimates of
the accretion rate and duration of the event provided one knows the
radius where the disk forms. This includes most of the models in Table
1. One must distinguish however, three classes of solutions: 1) models
where the disk is being fed at a rate governed by viscous processes
outside the region where neutrino losses are important - white dwarfs
plus black holes and black holes plus helium stellar cores are
examples; 2) models where the characteristic time scale is the viscous
time scale for the disk in a region where neutrino losses
approximately balance dissipation - merging neutron stars, black holes
plus neutron stars, and to some extent, the collapsar model; and 3)
models where the accretion rate is not governed by disk viscosity, but
by other circumstances. The prime example of this last case is a
collapsar where the viscous time in the disk, at least for late times
and reasonable choices of $\alpha$, is short compared to the free fall
time of matter at a much larger radius which falls inward and feeds
the disk. The general collapsar model is actually a complicated case
that can lie on the boundary of all three classes.

Consider first the slowest accreting models, those of class 2.  The
viscous time scale is approximately $t_{visc} = r^2 / \nu$, where $\nu
= \alpha H^2 \Omega_K$. Thus we have $t_{visc} = \alpha^{-1}
(H/r)^{-2} \Omega_K^{-1}$. For an advection-dominated disk, we have $H
\sim r$, so $t_{visc} \sim \alpha^{-1} \Omega_K^{-1}$.  If we assume
that a mass $M_{disk}$ deposited at a given radius $r_{disk}$ is
accreted on the viscous time scale, we can estimate the accretion rate
as $\dot M \sim M_{disk} / t_{visc}$.  In general, we find {\sl for
advective disks where neutrino losses are negligible}
\begin{eqnarray}
\tau \sim 2.7 \alpha^{-1} M_1^{-1/2} r_{disk,9}^{3/2} {\rm \ \ s} \\
\dot M \sim 0.37 \alpha M_{disk} M_1^{1/2} r_{disk,9}^{-3/2} \msun
\s^{-1} \\
\end{eqnarray}
Assuming the radii for disk formation - approximate values are given
by just the Roche radius - one obtains the characteristics of merging
white dwarf black hole pairs and black hole helium core pairs given in
Table 1. These agree well with the detailed models (Fryer \& Woosley
1998; Fryer et al.  1998). The accretion rate spans a large range
because high mass white dwarfs (and compact helium cores) form their
disks at smaller radii than those with low mass.

Next we consider neutrino dominated disks. We assume that neutrino
cooling is so efficient that any energy dissipated in the disk is
quickly radiated away.  In other words, one assumes that neutrino
cooling will produce a thin disk (Shakura \& Sunyaev 1973) where
advection will be unimportant.  One can do this by simply substituting
the neutrino cooling rate ($\S$3) into the standard thin disk
equations.  There is a question of which of the two types of neutrino
cooling described above will be dominant.  By solving the thin disk
equations for each type of cooling in turn, one can then check to see
whether this should be the dominant cooling term in the resulting
disk.  Doing so, one finds that the thin disk should be quite dense,
and, as a result, cooling by pair capture on nucleons should dominate
over $\nu-\bar\nu$ annihilation. The thin disk solution is obtained by
equating the cooling rate by pair capture to the energy dissipation
rate per unit volume $(3/8 \pi) \dot M \Omega_K^2 / H$.  One then
finds:
\begin{eqnarray}
T = 1.3 \times 10^{11} \alpha^{0.2} M_1^{-0.2} R^{-0.3} \K \\
H = 1.7 \times 10^4 \alpha^{0.1} M_1^{0.9} R^{1.35} \cm \\
\rho = 1.2 \times 10^{14} \alpha^{-1.3} M_1^{-1.7} R^{-2.55} \dot M_1
\g \cm^{-3} \\
V = 5.6 \times 10^8 \alpha^{1.2} M_1^{-0.2} R^{0.2} \cm \s^{-1} \\
\Omega = \Omega_K = 2.0 \times 10^5 M_1^{-1} R^{-1.5} \s^{-1} \\
l = 4.4 \times 10^{15} M_1 R^{0.5} \cm^2 \s^{-1} \\
\tau = R/V = 2.6 \times 10^{-4} \alpha^{-1.2} M_1^{1.2} R^{0.8} \s
\end{eqnarray}
where $M_1$, $R$, and $\dot M_1$ are the mass of the accreting black
hole in solar masses, the radius in gravitational radii $G M_1 / c^2$
(4.43 km for a 3 M\sun \ black hole), and the mass accretion rate in
solar masses per second. Note that the pair capture rate is
proportional to $X_{nuc}$, which is taken to be unity. These equations
then provide the estimates given in Table 1 for the merging neutron
stars and neutron star black hole pairs.  Again, for an assumed disk
viscosity $\alpha = 0.1$ and black hole mass, $M_1$ = 3 M\sun, they
are not far from the detailed models (Ruffert \& Janka 1998; Eberl,
Ruffert, \& Janka 1998) provided a disk radius of 50 km is adopted.
This solution is shown in Fig. 1 for comparison with our calculations.

Finally we consider the collapsar. The angular momentum distribution
inside a collapsing massive stellar core is unknown. Such calculations
as have been done (Heger, Woosley, \& Langer 1998) indicate an
increasing value for $l$ as one moves out in mass, rising from a ${\rm
few} \times 10^{16}$ cm$^2$ s$^{-1}$ in the inner few solar masses to
about 10$^{17}$ cm$^2$ s$^{-1}$ in the outer regions. One expects
considerable variations from star to star.  This range of $l$
corresponds to disk radii of 50 - 250 km and viscous time scales
(neutrino dominated) of about 0.1 s. However, this is considerably
shorter than the free fall time for the mass that feeds the disk. For
a stellar radius that contains most of the mass of, e.g., a 10 M\sun \
helium core, $\sim 5 \times 10^9$ cm, the free fall time scale 1338
s/$\rho^{1/2}$ is several seconds, and this time scale grows longer as
the accretion continues. The collapsar is probably unable to produce a
jet during its first 5 - 10 s because the infalling matter along the
rotational axis sweeps any energy deposited into the hole. So a time
scale of $\sim$ 10 s seems reasonable for the collapsar. This is
consistent with detailed 2D models (MacFadyen \& Woosley 1998).

\section{ENERGY CONVERSION EFFICIENCIES}

According to the current paradigm (Rees \& M\'esz\'aros 1992, 1994,
and subsequent publications; Katz 1994; Sari \& Piran 1997a,b), a
gamma-ray burst is formed when a relativistic outflow, most likely a
jet, suffers collisions both internally and with the surrounding
medium, producing shocks.  Typical values needed for the jet are
relativistic $\Gamma$'s of $\gtaprx ~100$. The total mass and energy
of the jet depends upon its opening angle and the efficiency for
converting its energy to gamma-rays, but should be of order $10^{-6} -
10^{-5} \msun$ and $10^{50} - 10^{51} \erg$. Modern gamma-ray burst
models thus divide into two categories: those that treat the shock
interaction and production of gamma-rays and ``afterglows" in other
wavelengths (e.g., Waxman 1997) and those that discuss the energy
source itself. We are interested here only in the latter problem.

The main problem in all current models for the GRB energy source is
how to convert some fraction (albeit small) of the net accretion
energy into directed relativistic outflow. Two general mechanisms have
been proposed - neutrino annihilation and magnetohydrodynamical (MHD)
mechanisms of various kinds. The former is easier to understand and to
model and will be dealt with first.

\subsection {Efficiency for neutrino annihilation}

Table 2 shows that a highly variable fraction of the accretion energy
will be emitted as neutrinos. In general a higher accretion rate and a
low viscosity favor high efficiencies, $L_\nu/\dot M c^2$. A large
rotation rate for the black hole also enhances the efficiency. Most of
these neutrinos come from pair capture on nucleons and are thus solely
of the electron flavor. The smaller fraction of neutrinos from pair
annihilation will be a mixture of three flavors. Everywhere in the
vicinity of the black hole, but especially along the rotational axis,
neutrino will encounter neutrino with a large flux. Thus neutrino
annihilation, $\nu + \bar \nu \rightarrow e^+ + e^-$, will deposit
some fraction of the accretion energy in regions where the mass
density may be small. These same neutrinos will also drive a wind from
the disk by their interaction with electrons and nucleons. Since the
gravitational potential, neutrino temperatures, and fluxes are similar
to those for neutron stars, the semi-analytic formulae of Qian \&
Woosley (1996) should hold approximately. This suggests that the wind
from the disk will be negligible except in the case of neutron
star--black hole mergers. There it might amount to $\sim 0.01 \msun 
\s^{-1}$.

We neglect here the interaction of neutrinos with all other matter and
estimate the neutrino annihilation efficiency using a simple
approximation to the disk geometry. The neutrino luminosity is assumed
to be concentrated in the equatorial plane (a good approximation given
the high temperature sensitivity of the rates) and thus to have a
luminosity $4 \pi r H(r) \dot q(r) \erg \cm^{-1} \s^{-1}$ where $H(r)$
is the disk scale height and $\dot q(r)$ the neutrino luminosity per
unit volume evaluated for the temperature and density at radius
$r$. The neutrino annihilation at any point above the disk is
calculated following the method described in Ruffert \et (1997):
\begin{eqnarray}
l^+_{\nu \bar{\nu}}(\nu_i \bar{\nu_i})
=A_1 \sum_k \frac{\Delta L_{\nu_i}^k}{d_k^2} 
\sum_{k^{'}} \frac{\Delta L_{\nu_i}^{k^{'}}}{d_{k^{'}}^2}
[\langle \epsilon \rangle_{\nu_i} + 
\langle \epsilon \rangle_{\bar{\nu_i}}] (1-cos \theta)^2 
\nonumber \\
+ A_2 \sum_k \frac{\Delta L_{\nu_i}^k}{d_k^2} 
\sum_{k^{'}} \frac{\Delta L_{\nu_i}^{k^{'}}}{d_{k^{'}}^2}
\frac {\langle \epsilon \rangle_{\nu_i} + 
\langle \epsilon \rangle_{\bar{\nu_i}}}
{\langle \epsilon \rangle_{\nu_i} \langle \epsilon
\rangle_{\bar{\nu_i}}}(1-cos \theta)
\end{eqnarray}
where $A_1=\sigma_0 
(C_1+C_2)_{\nu_i \bar{\nu_i}}/(12 \pi^2 c^5 m_e^2)
\approx 1.7 \times 10^{-44} {\rm \, cm \, erg^{-2} \, s^{-1}}$ 
and $A_2=\sigma_0 (C_3)_{\nu_i \bar{\nu_i}}/(4 \pi^2 c)
\approx 1.6 \times 10^{-56} {\rm \, cm \, s^{-1}}$ 
are the neutrino cross-section constants for electron 
neutrinos (including geometrical factors).  We model the 
disk as a grid of cells in the plane with neutrino 
mean energy ($\langle \epsilon \rangle_{\nu_i}$) and 
luminosity ($\Delta L_{\nu_i}^k$).  For each pair of cells, 
and every point above the scale height of the disk $H(r)$, we calculate 
distance ($d_k$) from each cell to that point, and the angle 
($\theta$) at which the neutrinos and anti-neutrinos from the pair of 
cells interact.  The summation over all pairs of cells gives 
the energy density from pair production at that point.

Each of the neutrino energies and luminosities includes the effects of
general relativity as the neutrinos travel through the black hole's
potential well.  Those neutrinos which annihilate near the surface of
the black hole were emitted further out in the potential well and
actually gain energy before annihilation.  However, the pairs they
produce must then climb further out of the potential well, an effect
that almost exactly cancels out the energy gain.  In addition, most of
the annihilation energy is produced well above the black hole and
general relativity plays a very minor role ($<$5\%) in the total
fireball energy.  We have not modeled neutrino propagation through a
Kerr metric, nor have we modeled the bending of geodesics, but it is
unlikely that these effects will significantly alter our results.

For our models, most of the neutrino emission is due to electron
capture and the luminosity density from pair production ($l^+_{\nu
\bar{\nu}}(\nu_i \bar{\nu_i})$) is dominated by the annihilation of
electron neutrinos and anti-neutrinos.  Although we include the
contribution of the $\mu$ and $\tau$ neutrinos for completeness, they
make up less than 10\% of the total luminosity density.

Integrating over the distance above the plane and taking advantage 
of the cylindrical symmetry of our disk:
\begin{equation}
2 \pi r \int_0^{\infty} l^+_{\nu \bar{\nu}}(\nu_i \bar{\nu_i}) dz 
\end{equation}
demonstrates the strong focusing of the pair fireball (Fig. 9) For
most of the models, over half of the energy is injected at equatorial
radii $\lesssim2\times10^6$\,cm.  Further integrating the luminosity
over the equatorial radius gives the total neutrino annihilation
luminosities.  The net momentum of the pairs produced by this
annihilation is directed {\it outward} along the angular momentum axis,
and most of this energy will escape.  
Table 4 summarizes the neutrino annihilation energies
and efficiencies for all models studied.  The efficiency of energy
conversion from neutrinos to a electron/positron pair fireball is both
a function of the total neutrino emission and the distribution of that
emission.  For example, although the $\dot{M}=0.1 \msun \s^{-1},
\alpha=0.1, a=0$ model has a lower luminosity than its $\alpha=0.01$
counterpart, because its emission is more centrally concentrated, it
converts a larger fraction of the neutrinos into electron/positron
pairs and its resultant fireball is more energetic.

Note that the energy conversion efficiencies are extremely high for
the $\dot{M}=10 \msun \s^{-1}$ models.  For these models, our
assumption that the disk is optically thin to neutrinos no longer
holds (the neutrino mean free path is $\sim 1000-5000$\,cm whereas the
disk scale height is $10^5$\,cm).  For the non-rotating model, our
estimated luminosity exceeds that of Eberl, Ruffert, \& Janka (1998)
by nearly a factor of 4.  Most of this factor arises from the fact
that our models predict that nearly all of the neutrino energy is
produced by electron capture (increasing our efficiency by a factor of
3), and hence, is dominated by electron neutrinos and anti-neutrinos.
Although we have limited our neutrino luminosity by the blackbody
limit, overestimates in the neutrino temperature no doubt account for
the remaining 30-40\% differences.  Our estimates for the $\dot{M}=10
\msun \s^{-1}$ models should be taken as an upper limit for the
neutrino annihilation luminosity, and the actual luminosity could be
as much as a factor of 5 lower.

\subsection {Efficiencies for MHD energy extraction}

It is also possible to extract energy from either the disk or the
black hole by MHD processes (e.g., Narayan \et 1992; M\'esz\'aros \&
Rees 1997; Katz 1997).  These are based on the expectation that the
differential rotation of the disk will rapidly amplify preexisting
magnetic fields until they approach equipartition with the gas kinetic
energy.  Examples of MHD energy extraction mechanisms include a
relativistic wind or jet driven from the disk surface (Katz 1997;
M\'esz\'aros \& Rees 1997), Parker instability within the disk leading
to reconnection and flares (Narayan \et 1992), or the Blandford-Znajek
mechanism for extracting the spin energy of the black hole (Blandford
\& Znajek 1977).

Of these, perhaps the one for which it is easiest to estimate the
efficiency is the Blandford-Znajek (BZ) mechanism.  Other MHD
processes might have very different efficiencies, particularly since
unlike the others the BZ mechanism depends directly on the black hole
spin, but we give here the BZ--efficiency as a rough estimate.

All the evolutionary scenarios in Table 1 give black holes which are
rotating very fast. This is inherent in the fact that they accrete a
fraction of their mass from a disk. Typical values of the spin
parameter $a$ are 0.5 and we adopt that here as a representative
value. The magnetic field in the accreting plasma is harder to
estimate. Here we follow a common assumption that the field will rise
to some fraction, which we guess might be 10\% of its equipartition
value (i.e., an energy density 1\% of $\rho v^2$).  Table 3 gives
values of $\rho v^2$; typical values are $10^{30} \erg \cm^{-3}$
implying a field strength of $10^{14} - 10^{15}$ gauss. The BZ jet
luminosity is then
\begin{equation}
L_{\rm BZ} \ \approx \ 10^{50} \ a^2 \ \left({{B} \over {10^{15} {\rm
gauss}}}\right)^2 \ \ \erg \s^{-1}
\end{equation}
Entries for our various models are given in Table 6.  The estimated
luminosities are comparable to those for neutrino annihilation listed
in Table 5.  However, if the magnetic field reaches full equipartition
with the gas, rather than $1\%$ of $\rho v^2$ as we have assumed, the
BZ luminosities could be two orders of magnitudes larger than those
listed.

Note that to generate the equipartition field requires not only a
large kinetic energy density in the vicinity of the of the black hole
but that the disk make many revolutions. Table 3 gives the number of
windings a blob of accreting matter will experience in each of our
models, which is generally around $\alpha^{-1}$.  The number required
to generate the field depends upon the primordial field in the
accreting matter and the efficiency of instabilities in creating
radial variations in a magnetic field whose poloidal component is
being wound by rotation (Katz 1997). We suspect that 10 windings is a
gross lower bound.

\section {CONCLUSIONS AND IMPLICATIONS FOR GAMMA-RAY BURSTS}

We have explored accretion at very high rates, 0.01 to $10 \msun
\s^{-1}$ into stellar mass black holes.  The black holes were of both
the stationary (Schwarzschild) and rapidly rotating (Kerr)
varieties. For a disk viscosity $\alpha = 0.1$ and accretion rates
larger than $\sim 0.05 \msun \s^{-1}$, we find that a situation of
global balanced power comes to exist in the disk interior to about
$10^8 \cm$ where neutrino emission, chiefly by pair capture on
nucleons, approximately balances the energy dissipated by viscosity
(Table 2). For lower disk viscosity, the necessary accretion rate to
achieve balanced power is reduced (see $\alpha = 0.01$, $\dot M = 0.01
\msun \s^{-1}$, Table 2). Loss of appreciable energy to neutrinos
leads to a cooler, denser, and thinner disk than the purely advective
solution (Fig.  1).  The sudden onset of neutrino losses, which are
very sensitive to the temperature and density, appreciably thins the
disk and may help to provide favorable geometry for jet acceleration.

Temperatures in the inner disk range from 2 MeV to 13 MeV in the models
studied and densities from $10^9$ to $10^{12} \g \cm^{-3}$ (Table
3). Only for the most extreme accretion rates, $\dot M \gtaprx ~1
\msun \s^{-1}$, does the disk start to become optically thick or the
neutrino emission blackbody limited. Thus one can calculate the
neutrino losses and the efficiency for neutrino annihilation using a
very simple approach.

We found that the efficiency for converting accreted mass energy, $\dot
M c^2$, into neutrinos is highly variable. Greater efficiency is
favored by low viscosity and high accretion rate, but in no case is the
full theoretical limit reached. In some cases the efficiency is very
low (e.g., 0.08\% for $\alpha = 0.1$ and $\dot M = 0.01 \msun
\s^{-1}$. The efficiency is also smaller if the black hole mass is
larger (Table 2).

We also calculated the efficiency for neutrino annihilation for all of
our models and found a very large range (Table 4). This range is a
consequence of the large fraction of accretion energy advected into
the hole for accretion rates under 0.05 M\sun \ s$^{-1}$, which
reduces the neutrino luminosity, and a reduction in mean neutrino
energy for the cooler, slower accreting disks, which reduces the cross
section for neutrino annihilation.  For a Schwarzschild black hole,
changing the accretion rate by one order of magnitude from 0.01 M\sun
\ s$^{-1}$ to 0.1 M\sun \ s$^{-1}$ changes the energy deposition by
neutrino annihilation by five orders of magnitude.  Above 0.1 M\sun \
s$^{-1}$ the efficiency for neutrino annihilation continues to
increase reaching at least several per cent for accretion rates of 5
M\sun \ s$^{-1}$ on a rapidly rotating black hole. It is impossible to
produce a bright GRB (of say 10$^{49}$ ergs which, with a beaming
factor of 100, might resemble a 10$^{51}$ erg burst) by neutrino
annihilation in any model in which the accretion rate is less than a
${\rm few} \times 10^{-2}$ M\sun \ s$^{-1}$. On the other hand
accretion rates over 0.1 M\sun \ s$^{-1}$, however they may be
realized, can give very energetic events (Tables 4 and 5).

With the energy densities calculated for our inner disks we also
estimate the efficiency for jet acceleration by the Blandford-Znajek
(BZ) process using a representative value for the spin parameter (0.5)
and a guess regarding the the magnetic energy density (1\% of $\rho
v^2$ in the inner disk). Typical total GRB energies were around
10$^{50}$ erg (times the beaming factor).  These estimates are very
crude; in particular, the field energy density in the disk could be
1\%, 100\%, or 0\% of equipartition.  We give these estimates only to
show that the kinetic energy density in the disk is sufficient to
anchor an adequate field to power a bright (beamed) GRB.

However, we note that these large fields might require many orbits of
the disk to generate. The number of disk windings is small unless the
viscosity is initially low (Table 3). One can envision a disk that has
essentially zero viscosity until the field is wound up to a
significant fraction of equipartition.  Then the estimates of field
strength in the inner disk (Table 6) would be higher. We see two
solutions to the GRB problem emerging - high viscosity disks that make
bursts by neutrino annihilation, and low viscosity disks that extract
rotational energy from the black hole by MHD processes. Our
understanding of the disk physics is inadequate to distinguish these at
the present time. We have concentrated on neutrino annihilation here
simply because it is easier to calculate.

The implications of our work for GRB models are best discussed on a
case by case basis (see Tables 5 and 6). The event rates in Table 5
are representative and are taken from the forthcoming paper by Fryer,
Woosley, \& Hartmann (1998).

\subsection{Merging Neutron Stars and Black Hole Neutron Star Pairs}

Our calculations here qualitatively confirm the more accurate, but
specialized work of Ruffert \& Janka (1998) and Eberl, Ruffert, \& Janka
(1998). Given that a disk forms at $\sim$50 km, eq. 5.9 gives the correct
approximate accretion time scale, 0.1 s, provided $\alpha$ = 0.1 and the
black hole mass is 3 M\sun. For other values of $\alpha$ and black hole
mass, our results suggest how their results might be scaled. Our
neutrino luminosity, about 10$^{53}$ erg s$^{-1}$, neutrino annihilation
rate, $5 \times 10^{50}$ erg s$^{-1}$, and total energy available for the
burst, $5 \times 10^{49}$ erg, agree with the detailed calculations of
merging neutron stars.

For neutron star plus black hole mergers, our calculations suffer from
the disk becoming optically thick. The time scale for accretion, still
0.1 s, agrees well, and for the larger disk mass (Table 1), it is
clear that greater energies and efficiencies will be developed. The
entries in Table 5 come from logarithmic interpolation in Table 4 to
obtain the neutrino annihilation luminosity for an accretion rate of 5
M\sun \ s$^{-1}$, then arbitrary division by 4 to correct for the
decrease in neutrino luminosity and cooling of neutrino energy one
expects for a (optically) thick disk. These numbers are then
consistent with Eberl, Ruffert, \& Janka (1998).

The chief new result of our calculations here is an estimate of the
effect of using Kerr geometry. The 3D numerical calculations of Eberl,
Ruffert \& Janka (1998) were all carried out for a non-rotating black
hole. Here we see that a Kerr parameter of only 0.5 increases the
efficiency of neutrino (emission and) annihilation by a factor of 4 to
6. This degree of rotation is reasonable for a black hole made by
merging neutron stars, and, depending on its formation process, might
be reasonable for a black hole merging with a neutron star.

Of all the models in Table 5, only the merging neutron stars and black
hole--neutron star pairs have a characteristic time scale much less
than one second. They are thus the only model capable of explaining
the abundant subclass of short (average duration 0.3 s) hard GRBs
(Fishman \& Meegan 1995). We have no information on the optical
counterparts of these short bursts, and do not know if they are
associated with star formation.

Because the initial conditions are so uniform, especially for merging
neutron stars, one might expect that the jet has a nearly unique
energy and duration. GRB diversity in this case would solely reflect the
variable characteristics of the interstellar medium in which the burst was
embedded.

\subsection {Collapsars}

Here our results are more sensitive to uncertain parameters and a
diversity of outcomes is possible. In addition to the uncertain disk
viscosity which plagues all accretion powered models, the accretion
rate is particularly sensitive to the angular momentum distribution of
the accreting star.  This is poorly known and likely to be highly
variable both within a star and from star to star.  The radius where
the disk forms varies as $l^2$ (eq. 5.8) and the accretion rate as an
even higher power.  Fortunately, if the disk viscosity is adequately
high, and if a steady state disk forms, the accretion rate may be
determined by simpler physics -- the rate at which matter is delivered
to the accretion disk by the collapsing star.  Our estimate comes from
taking the free fall time of the star divided into that fraction of
the star's mass which forms a bound disk.  We estimate 10 s for the
former and perhaps a few solar masses for the latter.  The frequency
of events is also very uncertain.  In Table 5 we have taken from 0.5\%
to 5\% of the supernova rate.

The formation of a bound disk requires dissipation.  The disk of a
collapsar may be unique in nature in that it is assembled from
optically thick matter falling in, essentially freely from
infinity.  In the absence of neutrino losses and photodisintegration
(which only occur inside a few hundred km), the net binding energy of
the disk must be carried away by outflowing matter.  This can be a
considerable amount of energy.  For example, a 1 M\sun disk forming at
250km ($l \simeq 10^{17} \cm^2 \s^{-1}$) must release a binding 
energy of $3 \times 10^{52} \erg$.  Half of this is rotational energy
and most of the rest is internal energy.  But if even a small fraction
goes into outflows, they could power an energetic supernova like SN
1998bw (Woosley, Eastman, \& Schmidt 1998).  Because a lot of the
matter that flows out will have been at small radii and very hot, its
composition may be rich in $^{56}$Ni.  Our simple steady state
calculations here cannot model the multi-dimensional physics of disk
formation and bi-directional flow.  Two dimensional studies are in
progress (MacFadyen \& Woosley 1998).

But provided enough dissipation occurs, by mass ejection, viscous
interaction, photodisintegration, and neutrino losses, something
resembling a steady state accretion disk will form. As the accretion
proceeds the black hole will be spun up. Our results show (Table 4)
that the energy available for jet production from neutrino
annihilation is highly variable ranging from essentially zero, if the
accretion rate is 0.01 M\sun \ s$^{-1}$ or less, to over 10$^{52}$ erg
if the black hole rotates rapidly and the accretion rate exceeds 0.1
M\sun \ s$^{-1}$ for 10 s (Table 5). This makes collapsars potentially
the most energetic and most frequent of all GRB models.  If the
beaming factor is $\sim$100, as preliminary calculations by MacFadyen
\& Woosley (1998) indicate, GRBs of equivalent energy up to 10$^{54}$
erg might be explained.

It should be noted that in this and all models with long time scales,
the energy available for a GRB is the GRB duration times the jet
luminosity - not necessarily the total energies in Tables 5 and
6.  This is another reason why only merging neutron stars and black
holes can make energetic GRBs shorter than 1 s.

\subsection {White Dwarfs Plus Black Holes}

For these models the accretion rate is expected to be lower. The Roche
radius gives a range of disk sizes and, for a given disk viscosity and
white dwarf mass, this sets the range of accretion rate (Table 1). For
the lower values, the energy from neutrino annihilation is
negligible. For the higher mass white dwarfs which give accretion
rates of 0.07 M\sun \ s$^{-1}$ for 15 s, the energy is dramatically
greater. If the black hole rotates, the yield is further increased to
perhaps 10$^{50}$ erg.  We expect the beaming factor to be less here
than in the collapsar, but bursts of up to 10$^{51}$ erg might be
explained, more if the Blandford-Znajek mechanism is effective (Table
6). However, unless these events occur more frequently than black
hole--neutron star mergers, they are not likely to be the leading
cause of GRBs.

\subsection {Black Holes and Common Envelope Evolution}

As with the collapsar which it resembles, this is a model where
critical parameters remain poorly determined. The three dimensional
evolution of a black hole merging with a helium core (or helium star)
of comparable or greater mass has not been studied. One expects a
great deal of the accretion to occur along the poles. The radius at
which the helium disk forms is poorly known. {\sl If} it is as small
as 10$^4$ km for at least 1 M\sun \ of helium, accretion rates of 0.1
M\sun \ s$^{-1}$ may be realized. If the black hole further has a Kerr
parameter of 0.5, the neutrino annihilation energy would be
$\sim 10^{51} \erg$. If the rotation of the black hole becomes even
faster as a consequence of the merger, the energy could be increased
still more. Because of the large mass and angular momentum in the
merger, the channeling of the jet may result in tighter beaming,
comparable to the collapsar and more focused than merging compact
objects.

\acknowledgments This research has been supported by a cooperative
grant between the MPA, Garching, and UCSC from the NSF and DAAD (NSF
INT 9726315 and DAAD 315) and at UCSC by NASA (NAG5-2843 and MIT SC
A292701), and the NSF (AST-97-31569).  We acknowledge many helpful
conversations on the subject of gamma-ray bursts with Andrew
MacFadyen, Alex Heger, Max Ruffert, Thomas Janka, and Eli Waxman, and
thank Jonathan Katz, Peter M\'esz\'aros, Bohdan Paczy\'nski, Tsvi
Piran, and Martin Rees for useful comments.

\newpage

\begin{table*}[t]
\begin{center}
\centerline {TABLE~1}
\vskip 8pt
\begin{tabular}{lllllll}
\hline\hline
Model & M$_{\rm accrete}$& Radius& $\dot {\rm M}$  & j
& Duration & Gravity \\
      & (M\sun$^a$)      & (km)  & (M\sun/s) & (10$^{16}$ cm$^2$
s$^{-1}$) & (sec) & Waves   \\
\hline
n*+n*      & 0.1  & 50                   &   1  &  4       & 0.1       & Yes
\\
n*+BH      & 0.5  & 50                   &   5  &  4       & 0.1       &
Yes \\
Collapsar  & 2    & 50 - 250             & 0.1  & 5 - 10   & 10-20     & No
\\
BH+WD      & 1    & (1-5)$\times 10^4$   & 0.01 - 0.07 & 50 - 150 & 15 -
150 & No \\
BH+He core & 2    & (1- 10)$\times 10^4$ & 0.01 - 0.1  & 50 - 200 & 15 -
500 & No \\
\hline\hline
\end{tabular}
\end{center}
$^a$ Masses are for accretion through a disk. The total accretion
rate, e.g. in the collapsar and He core models, is greater because of
mass infall along the poles. The assumed mass of the black hole is 3
M\sun \ in all cases and the disk viscosity, $\alpha = 0.1$.
\end{table*}

\clearpage

\newpage

\begin{table*}[t]
\begin{center}
\centerline {TABLE~2}
\vskip 8pt
\begin{tabular}{llll|lllllll}
\hline\hline
\multicolumn{4}{c|}{Model}&\multicolumn{7}{c}{Luminosities ($10^{51}
\erg/\s$)} \\
$\dot M$ & $\alpha$ & $a$ & $M$ & $L_\nu$ & $L_\nu/\dot M c^2$ &
$L_{\nu, ann}$ & $L_{\nu, URCA}$ & $L_{photo}$ & $L_{diss}$ &
$L_{ent}$ \\
\hline
0.01 & 0.1 & 0 & 3 & 0.015 & 8.6e-4 & 3.4e-3 & 0.012 & 0.200 & 0.859 &
0.644 \\
0.01 & 0.03 & 0 & 3 & 0.089 & 5.0e-3 & 0.012 & 0.078 & 0.200 & 0.738 &
0.450 \\
0.01 & 0.01 & 0 & 3 & 0.650 & 0.036 & 0.012 & 0.638 & 0.200 & 0.724 &
-0.124 \\ 
0.01 & 0.1 & 0.5 & 3 & 0.036 & 2.0e-3 & 9.9e-3 & 0.026 & 0.200 & 1.24
& 1.01 \\
0.01 & 0.01 & 0 & 10 & 0.049 & 2.7e-3 & 9.3e-3 & 0.040 & 0.200 & 0.431
& 0.181 \\
0.05 & 0.1 & 0.5 & 3 & 1.65 & 0.018 & 0.36 & 1.34 & 1.00 & 6.48 & 3.86
\\
0.1 & 0.1  & 0   & 3 & 3.35 & 0.019 & 0.39 & 2.96 & 2.00 & 9.18 & 3.88
\\ 
0.1 & 0.03 & 0 & 3 & 6.96 & 0.039 & 0.076  & 6.89 & 1.99 & 7.94 &
-1.01 \\
0.1 & 0.01 & 0 & 3 & 6.15 & 0.034 & 7.7e-3 & 6.14 & 2.00 & 7.44 &
-0.71 \\ 
0.1 & 0.1 & 0.5 & 3 & 8.03 & 0.045 & 1.03 & 7.00 & 2.00 & 12.96 & 3.02
\\
0.1 & 0.1 & 0.95 & 3 & 46.4 & 0.26 & 2.61 & 43.8 & 1.99 & 36.8 & -11.5
\\
0.1 & 0.1 & 0.95 & 6 & 26.2 & 0.15 & 7.24 & 19.0 & 2.00 & 31.1 & 3.20
\\ 
1.0 & 0.1 & 0 & 3 & 86.3 & 0.048 & 0.66 & 85.7 & 19.9 & 90.9 & -14.9
\\ 
1.0 & 0.1 & 0.5 & 3 & 142. & 0.078 & 1.53 & 140. & 19.9 & 142. & -19.6
\\
10.0 & 0.1 & 0 & 3 & 781. & 0.043 & 53.3 & 728. & 200. & 836. &
-135. \\ 
10.0 & 0.1 & 0.5 & 3 & 1285. & 0.071 & 174. & 1111. & 200. & 1263. &
-211. \\
\hline\hline
\end{tabular}
\end{center}
\end{table*}

\newpage

\begin{table*}[t]
\begin{center}
\centerline {TABLE~3}
\vskip 8pt
\begin{tabular}{llll|llll}
\hline\hline
\multicolumn{4}{c|}{Model}&\multicolumn{4}{c}{Values at horizon}\\
$\dot M$ & $\alpha$ & $a$ & $M$ & $\rho$ & $T$ &
$\rho V^2$ & $N_{wind}$\\
$\msun \s^{-1}$ & & & $\msun$ & $10^{10} \g \cm^{-3}$ & $10^{10}$ K &
$10^{30} \erg \cm^{-3}$ & \\
\hline
0.01 & 0.1  & 0    & 3  & 0.060   & 3.90 & 0.53  & 7.1  \\
0.01 & 0.03 & 0    & 3  & 0.084   & 3.66 & 0.75  & 23.4 \\
0.01 & 0.01 & 0    & 3  & 0.23    & 1.91 & 2.04  & 90.2 \\
0.01 & 0.1  & 0.5  & 3  & 0.080   & 4.61 & 0.70  & 7.6  \\
0.01 & 0.01 & 0    & 10 & 9.32e-3 & 1.96 & 0.083 & 19.7 \\
0.05 & 0.1  & 0.5  & 3  & 0.43    & 6.67 & 3.84  & 9.8  \\
0.1  & 0.1  & 0    & 3  & 0.69    & 6.46 & 6.14  & 9.6  \\
0.1  & 0.03 & 0    & 3  & 1.67    & 3.47 & 15.0  & 31.2 \\
0.1  & 0.01 & 0    & 3  & 2.10    & 2.04 & 18.8  & 87.9 \\
0.1  & 0.1  & 0.5  & 3  & 1.07    & 7.42 & 9.45  & 10.2 \\
0.1  & 0.1  & 0.95 & 3  & 12.0    & 5.83 & 98.7  & 14.0 \\
0.1  & 0.1  & 0.95 & 6  & 0.88    & 9.04 & 7.46  & 10.2 \\
1.0  & 0.1  & 0    & 3  & 12.6    & 5.71 & 113.  & 9.9  \\
1.0  & 0.1  & 0.5  & 3  & 20.1    & 6.35 & 178.  & 11.5 \\
10.0 & 0.1  & 0    & 3  & 86.6    & 11.4 & 774.  & 6.6  \\
10.0 & 0.1  & 0.5  & 3  & 135.    & 13.1 & 1197. & 7.4  \\
\hline\hline
\end{tabular}
\end{center}
\end{table*}

\newpage

\begin{table*}[t]
\begin{center}
\centerline {TABLE~4}
\vskip 8pt
\begin{tabular}{llll|lll}
\hline\hline
$\dot M$           & $\alpha$ & $a$ & M & L$_{\nu}$                &
L$_{\nu \bar \nu}$         & efficiency \\
(M\sun \ s$^{-1}$) &   &  & $\msun$   & (10$^{51}$ erg s$^{-1}$) &
(10$^{51}$ erg s$^{-1}$)   &      (\%)       \\
\hline
0.01       &  0.1   &  0   & 3  & 0.015 & $3.9\times10^{-8}$ & 0.0003 \\
0.01       &  0.03  &  0   & 3  & 0.089 &                    &        \\
0.01       &  0.01  &  0   & 3  & 0.650 & $9.0\times10^{-6}$ & 0.001  \\
0.01       &  0.1   & 0.5  & 3  & 0.036 & $5.9\times10^{-7}$ & 0.002  \\
0.01       &  0.01  &  0   & 10 & 0.049 & $6.4\times10^{-9}$ & $10^{-5}$ \\
0.05       &  0.1   & 0.5  & 3  & 1.65  & $1.8\times10^{-3}$ & 0.11  \\
0.1        &  0.1   &  0   & 3  & 3.35  & $3.0\times10^{-3}$ & 0.09  \\
0.1        &  0.03  &  0   & 3  & 6.96  &                    &       \\
0.1        &  0.01  &  0   & 3  & 6.15  & $8.0\times10^{-4}$ & 0.01  \\
0.1        &  0.1   & 0.5  & 3  & 8.03  & 0.039              & 0.5   \\
0.1        &  0.1   & 0.95 & 3  & 46.4  & 2.0                & 4.2   \\
0.1        &  0.1   & 0.95 & 6  & 26.2  & 0.79               & 3.0   \\
1.0        &  0.1   &  0   & 3  & 86.3  & 0.56               & 0.6   \\
1.0        &  0.1   & 0.5  & 3  & 142   & 3.5                & 2.5   \\
10.0$^a$   &  0.1   &  0   & 3  & (781) &(200)               &(26)   \\
10.0$^a$   &  0.1   & 0.5  & 3  & (1280)&(820)               &(64)   \\
\hline\hline
\end{tabular}
\end{center}
$^a$ The assumption that the neutrinos are optically thin breaks 
down for accretion rates of 10 M\sun s$^{-1}$ and above.  The 
neutrino annihilation luminosities and energies listed for these high
accretion simulations are upper limits.
\end{table*}

\newpage

\begin{table*}[t]
\begin{center}
\centerline {TABLE~5}
\vskip 8pt
\begin{tabular}{llllll}
\hline\hline
Model & Duration & a & log L$_{\nu \bar \nu}$ & log E$_{\nu \bar \nu}$ &
Rate \\
      & (s)     &  & (erg s$^{-1}$) & (erg) &
(Myr$^{-1}$)$^a$ \\
\hline
n*+n*      & 0.1  &  0    & 50.7          & 49.7       &   0.1-5    \\
	   &      &  0.5  & 51.5          & 50.5       &            \\
n*+BH      & 0.1  &  0    & (52)$^b$      & (51)$^b$   &  0.1-50    \\
	   &      &  0.5  & (52.6)$^b$    & (51.6)$^b$ &            \\
Collapsar  & 10   &  0    & 48.5          & 49.5       &  $\lesssim 2000$ \\
	   &      &  0.95 & 51.3          & 52.3       &            \\
BH+WD      &15-150&  0    & 43.6 - 47.7   & 46-49      &  0.1-50    \\
	   &      &  0.5  & 44.8 - 48.8   & 47-50      &            \\
BH+He core &15-500&  0    & 43.6 - 48.5   & 46-50      &  1-1000    \\
	   &      &  0.5  & 44.8 - 49.6   & 47-51      &            \\
\hline\hline
\end{tabular}
\end{center}
$^a$This rate assumes a supernova rate of 0.02 yr$^{-1}$.

$^b$Rough estimate owing to optically thick disk.
\end{table*}

\clearpage

\newpage

\begin{table*}[t]
\begin{center}
\centerline {TABLE~6}
\vskip 8pt
\begin{tabular}{lllll}
\hline\hline
Model$^a$ & Duration & B$_{15}$ & log L$_{\rm BZ}$ & log E$_{\rm BZ}$ \\
      & (s)          & (10$^{15}$ gauss) & (erg s$^{-1}$) & (erg)     \\
\hline
n*+n*      & 0.1  & 10   & 51 & 50  \\
n*+BH      & 0.1  & 10   & 51 & 50  \\
Collapsar  & 10   &  1   & 49 & 51  \\
BH+WD      &15-150&  1   & 49 & 50  \\
BH+He core &15-500& 0.3  & 48 & 51  \\
\hline\hline
\end{tabular}
\end{center}
$^a$Assuming a=0.5 in all cases
\end{table*}

\clearpage

\figcaption{A neutrino-cooled accretion disk solution
(solid line) for an accretion rate $\dot M = 1 \msun \s^{-1}$, black
hole spin $a=0$, viscosity parameter $\alpha = 0.1$, and black hole
mass $M = 3 \msun$.  The six panels show the density $\rho$,
temperature $T$, radial velocity $V$, disk height $H$, angular
velocity $\Omega$, and specific angular momentum $l$.  Also shown for
comparison purposes are a pure advection solution with no cooling
(dotted line) and an analytic thin disk solution where URCA-process
neutrino cooling balances viscous dissipation (dashed line).}

\figcaption{Additional parameters for the solution shown
in Fig. 1.  Note that here the dashed and dotted lines have
different meanings than in Fig. 1.  The upper left panel shows the
cooling rate due to neutrinos and photodisintegration (solid line),
the heating rate due to viscous dissipation (dotted line), and the
rate at which energy is added to the entropy of the gas (dashed line).
The upper right panel shows $f = 1 - Q_{cool}/Q_{diss}$; the feature
around $\log r = 6$ occurs because $Q_{diss}$ briefly becomes negative
here.  The middle left panel shows the neutrino luminosity per unit
radius.  The middle right panel shows the cooling and accretion
time scales.  The two bottom panels show the pressure components: total
pressure (solid), gas pressure (dotted), pressure due to radiation and
pairs (short-dashed), and degeneracy pressure (long-dashed).}

\figcaption {Solutions for four accretion rates: $\dot M =
0.01$ (short-dashed), 0.1 (solid), 1 (dotted), and $10 \msun \s^{-1}$
(long-dashed).  All four have $a=0$, $\alpha = 0.1$, and $M = 3
\msun$.  The panels are as in Fig. 1, except for the two bottom panels
which show the neutrino luminosity per unit radius and $f$.}

\figcaption {Solutions for three values of the viscosity parameter
$\alpha = 0.1$ (solid), 0.03 (dotted) and 0.01 (dashed).  All three
solutions have $\dot M = 0.1 \msun \s^{-1}$, $a=0$, and $m = 3
\msun$.}

\figcaption {Same as Fig. 4, except that here the three solutions have
$\dot M = 0.01 \msun \s^{-1}$.}

\figcaption {Solutions for three values of the black hole
spin: $a = 0$ (solid), 0.5 (dashed), and 0.95 (dotted).  All three
solutions have $\dot M = 0.1 \msun \s^{-1}$, $\alpha = 0.1$, and $M =
3 \msun$.}

\figcaption {Same as Fig. 3, except that here the four
solutions all have $a=0.5$.}

\figcaption {The Kerr parameter $a$ as a function of the fractional
increase in the mass of a black hole accreting from a thin disk.}

\figcaption {(a) Pair annihilation luminosity per cm versus radius for
$\dot M=0.1 \msun \s^{-1}$ and $M_{\rm BH} = 3.0 \msun$.  In addition,
the total integrated luminosity out to each radius is plotted for the
same models.  These models are representative of the entire set, for
all of which over half of the energy is injected in a narrow beam with
equatorial radii $\lesssim 2\times10^6$\,cm.  (b) Pair annihilation
luminosity per cm versus distance along the angular momentum axis.
Again, the total integrated luminosity out to each scale height above
the plane is plotted for the same models.}

\newpage

\epsfysize=8.5truein \epsfbox{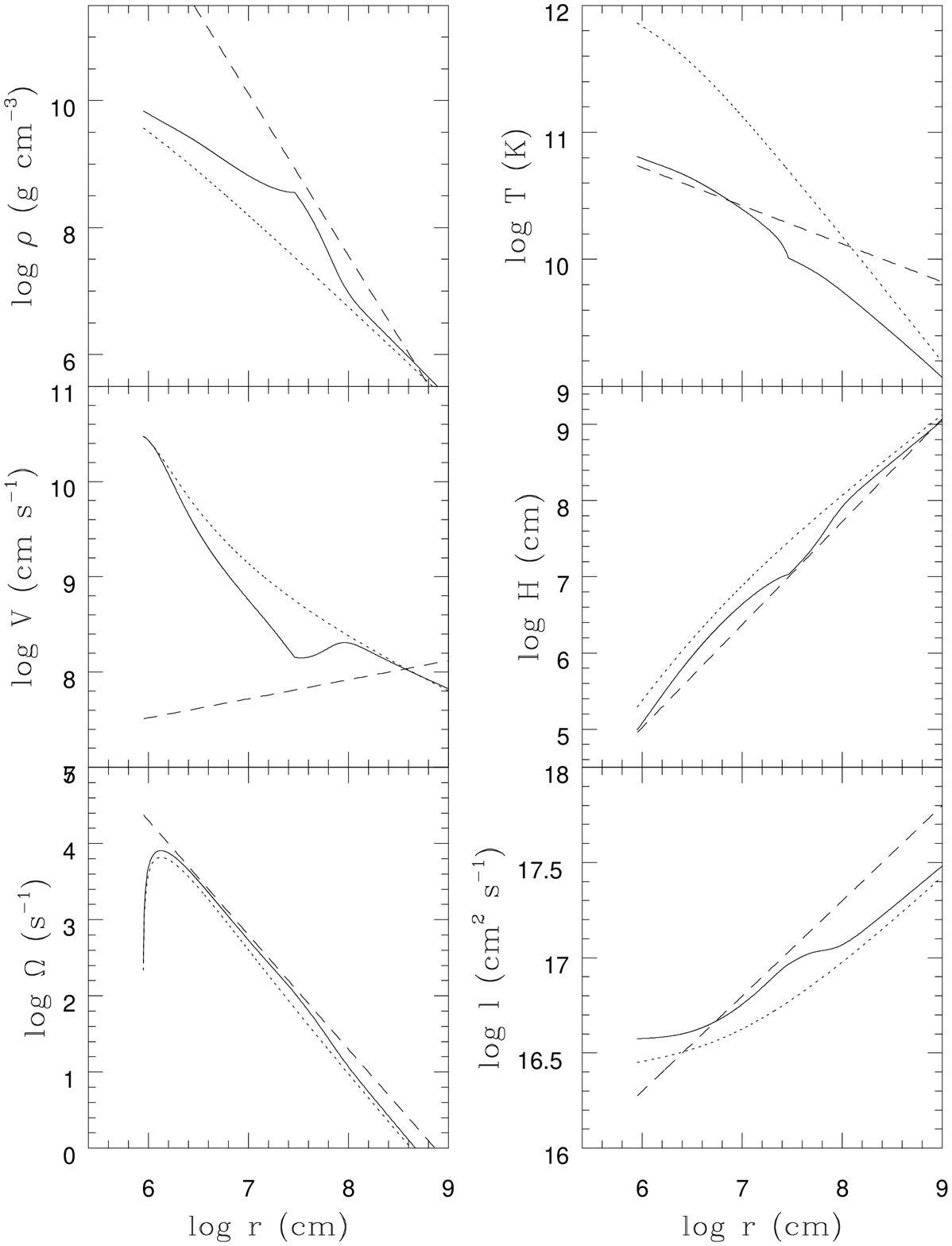}
\epsfysize=8.5truein \epsfbox{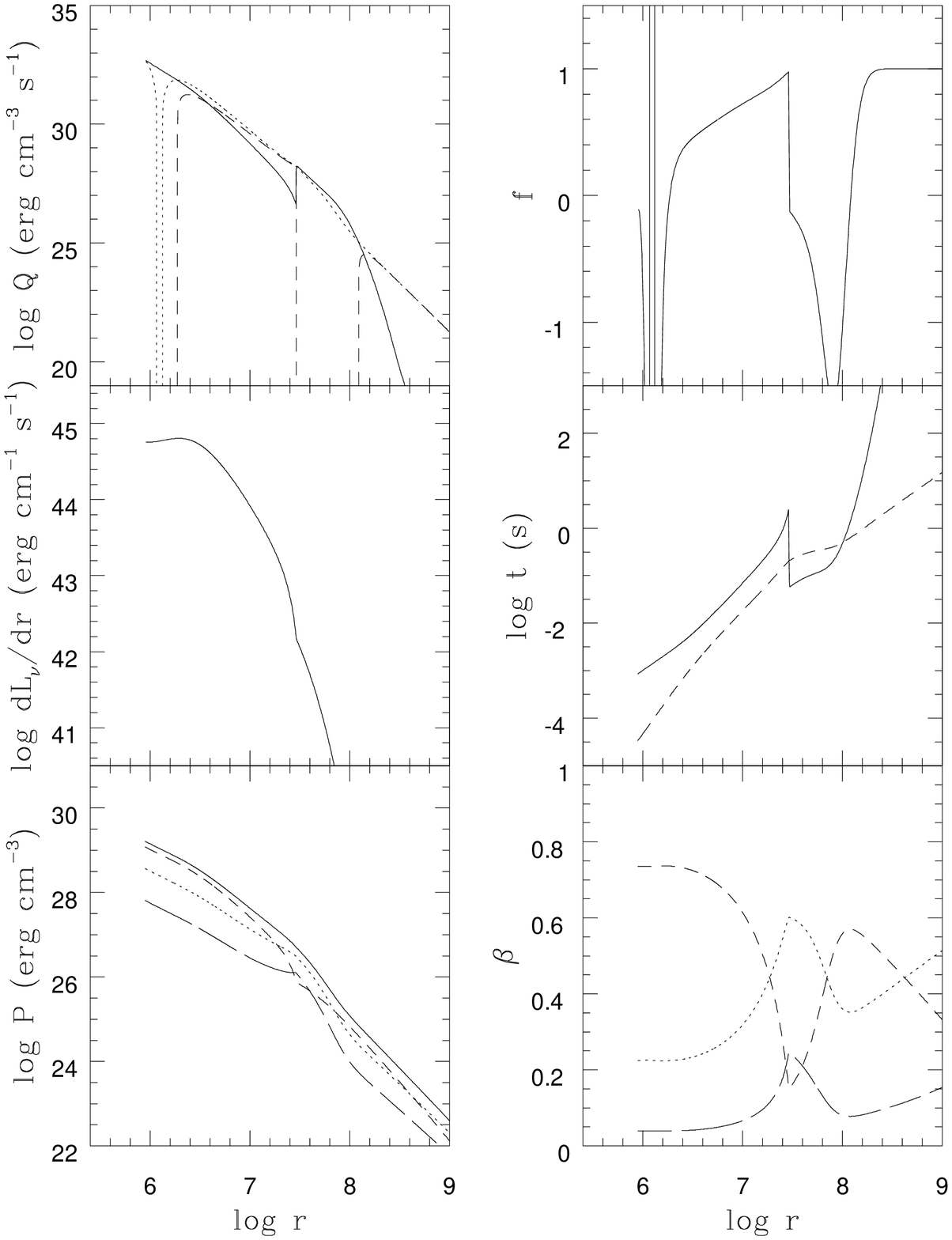}
\epsfysize=8.5truein \epsfbox{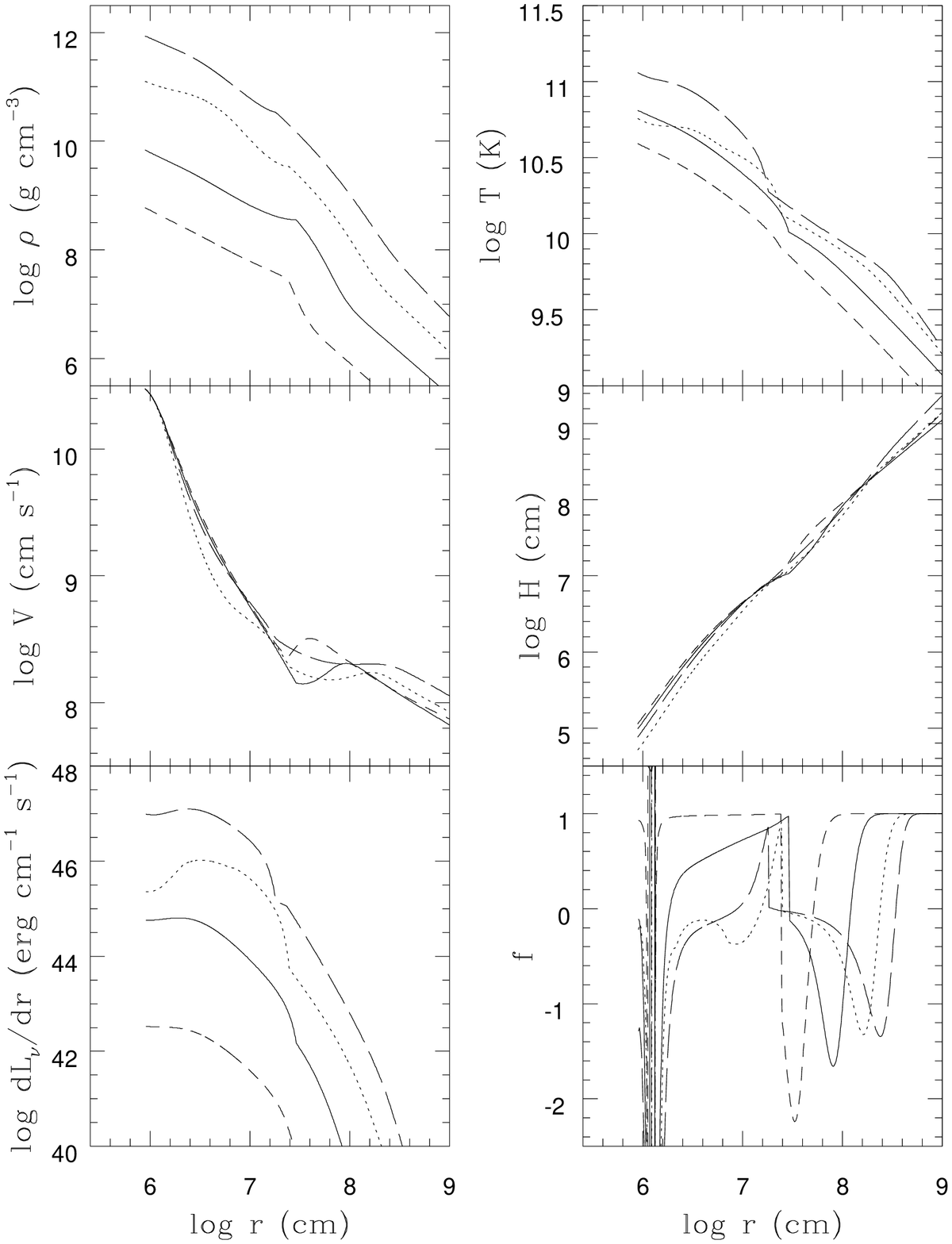}
\epsfysize=8.5truein \epsfbox{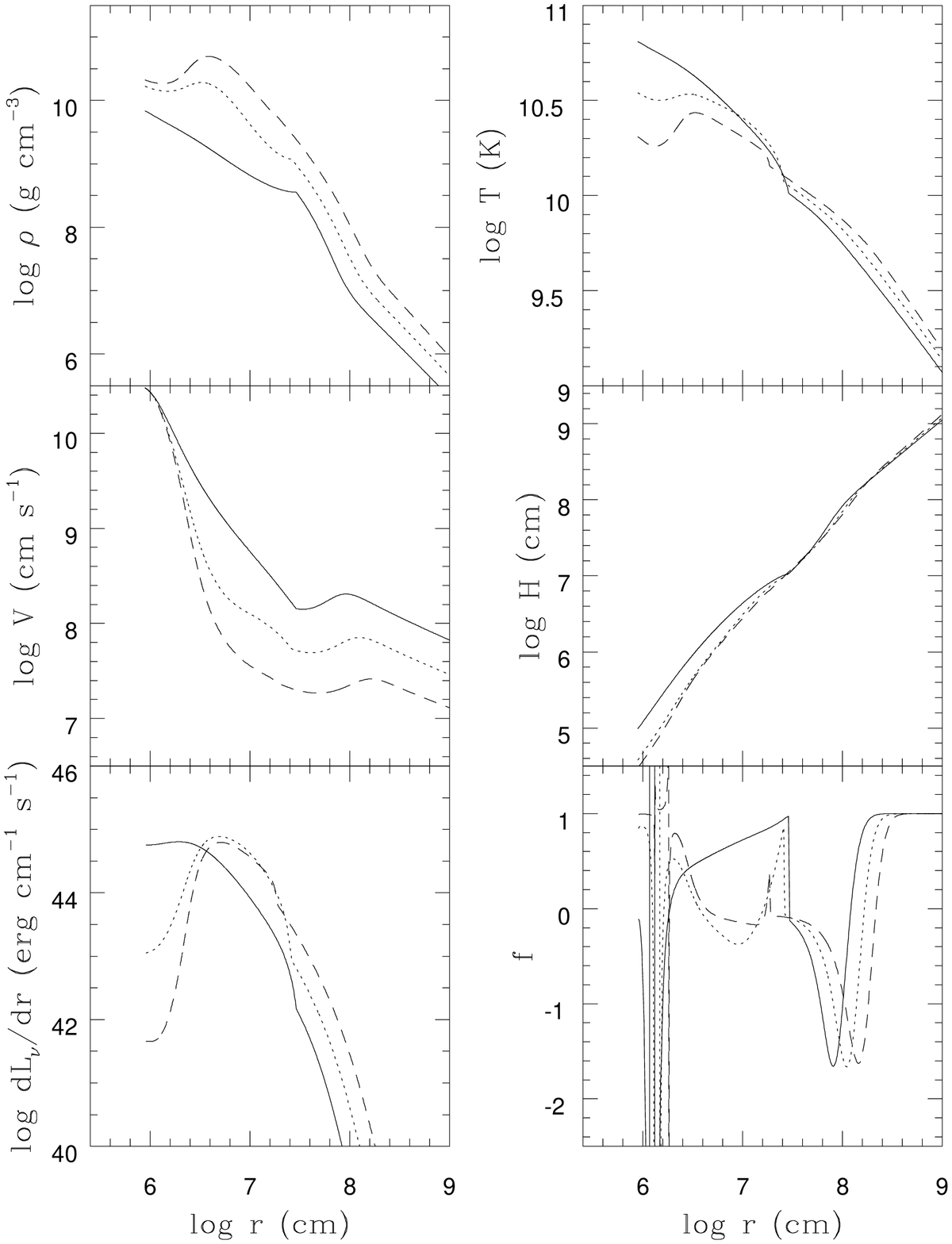}
\epsfysize=8.5truein \epsfbox{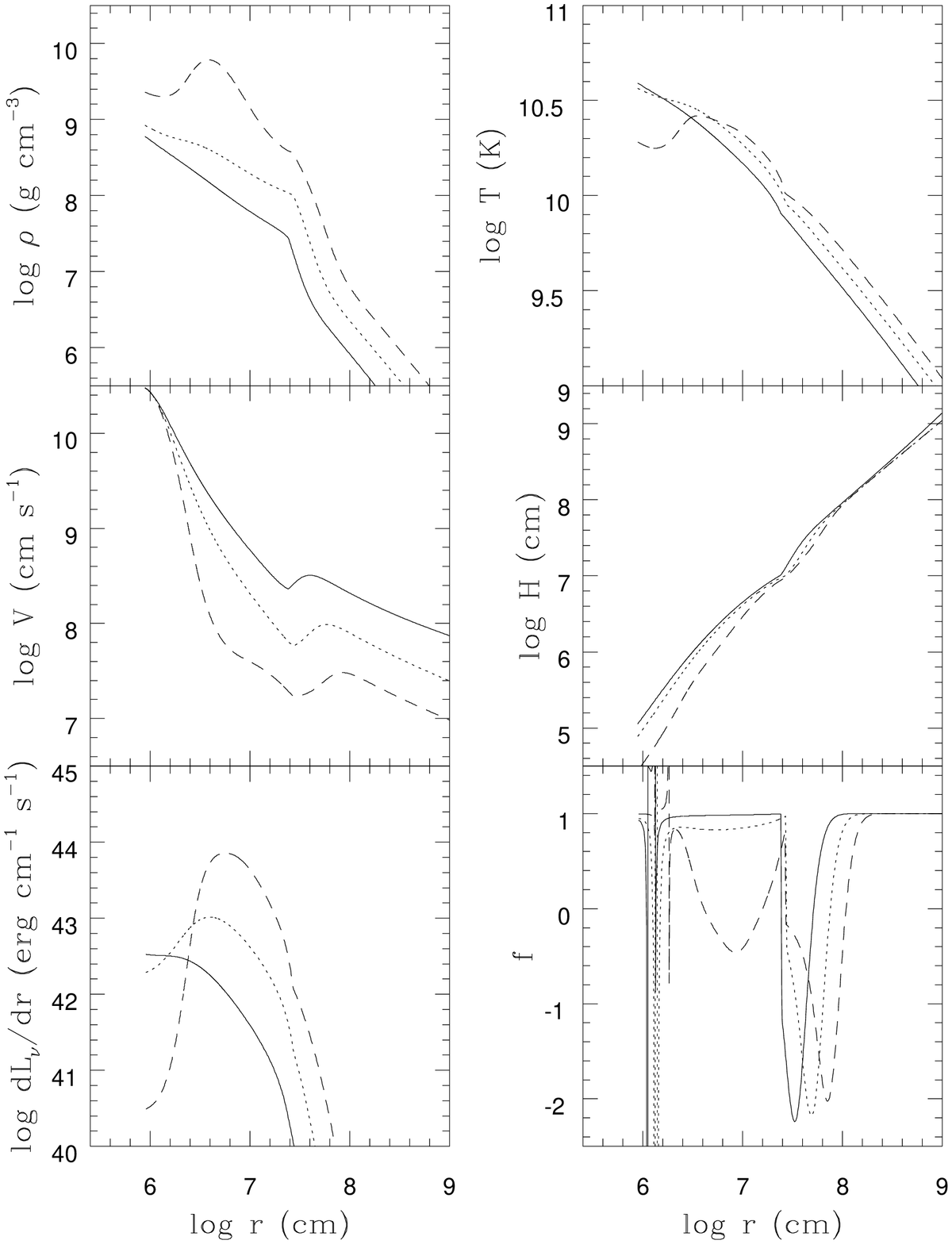}
\epsfysize=8.5truein \epsfbox{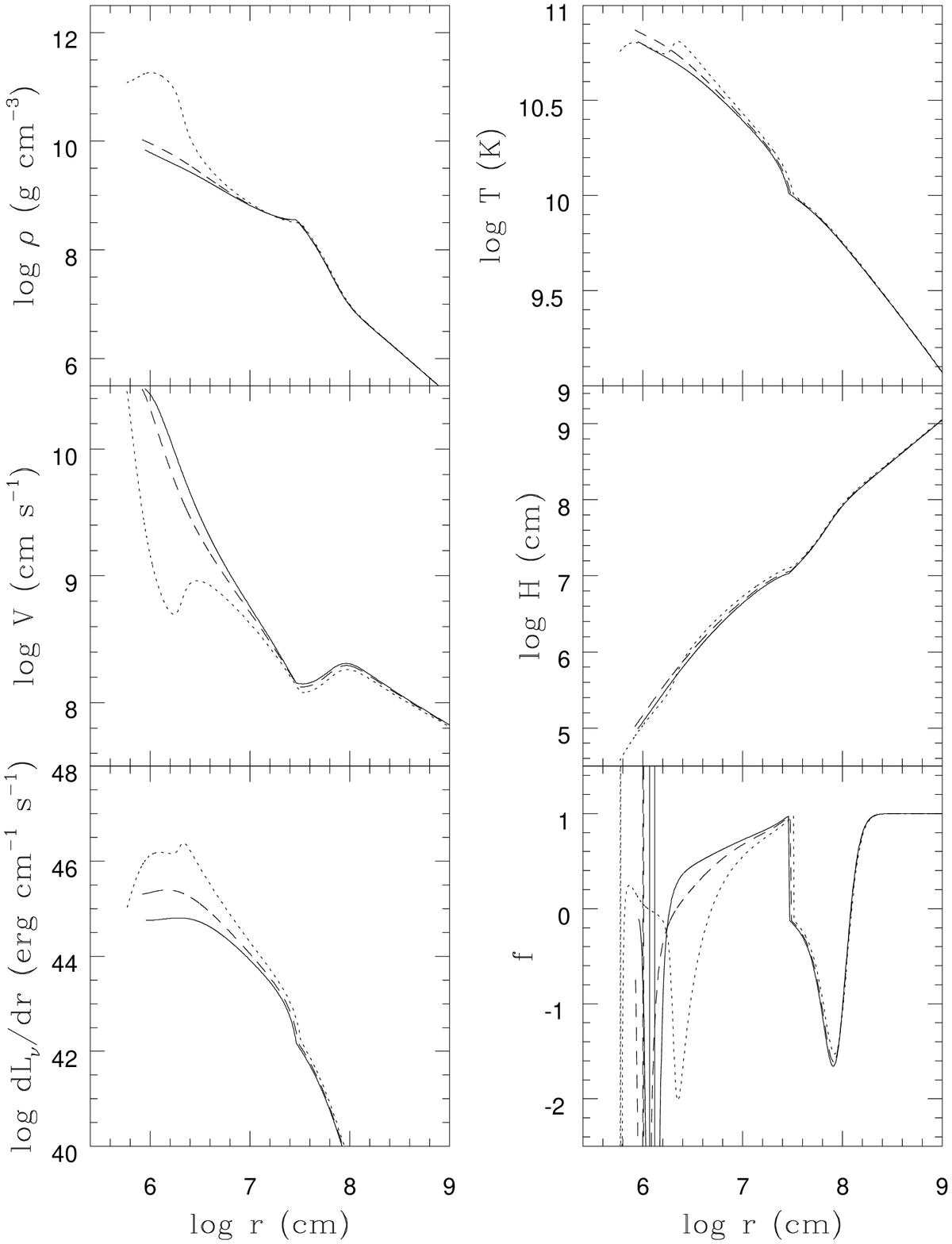}
\epsfysize=8.5truein \epsfbox{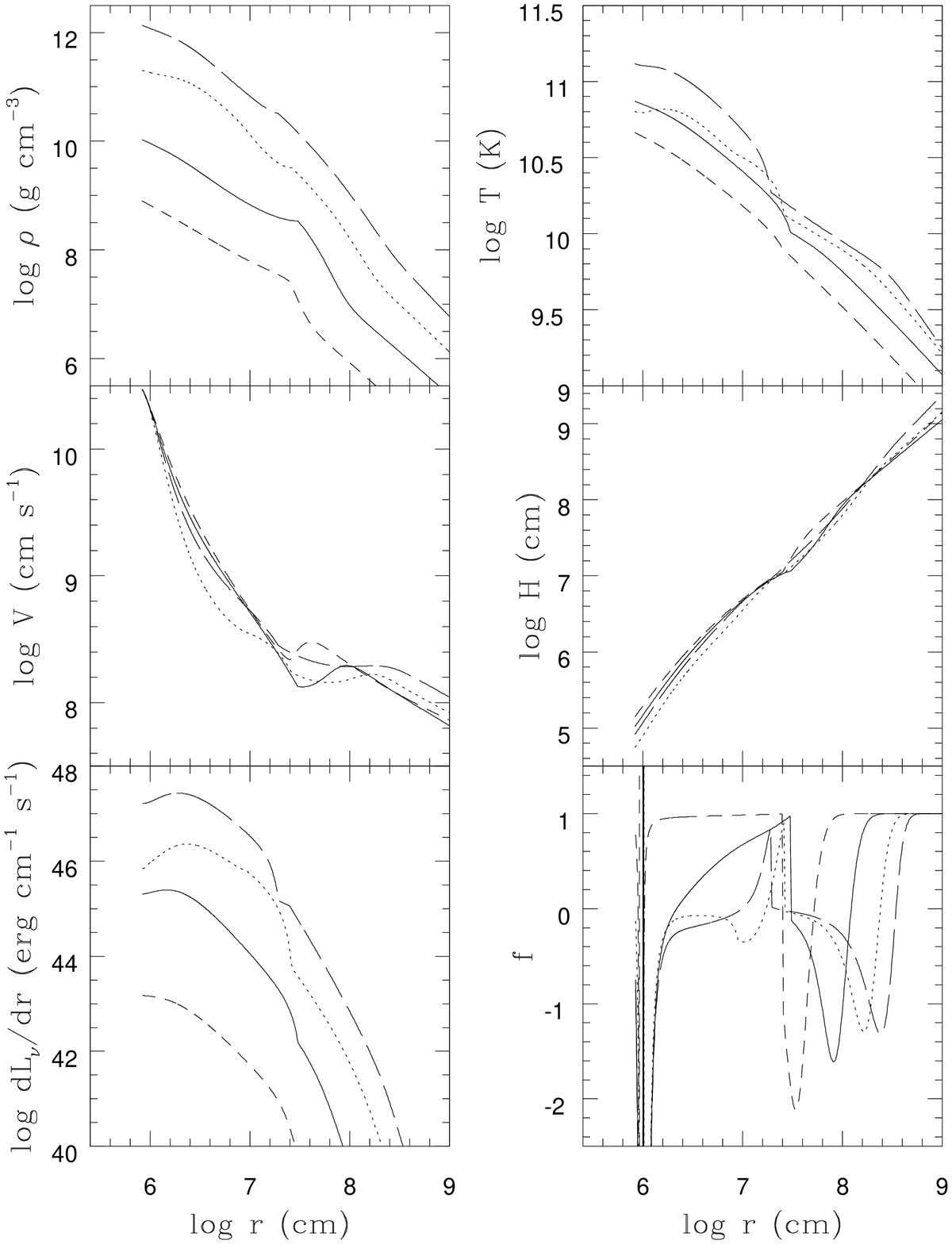}
\epsfysize=6.5truein \epsfbox{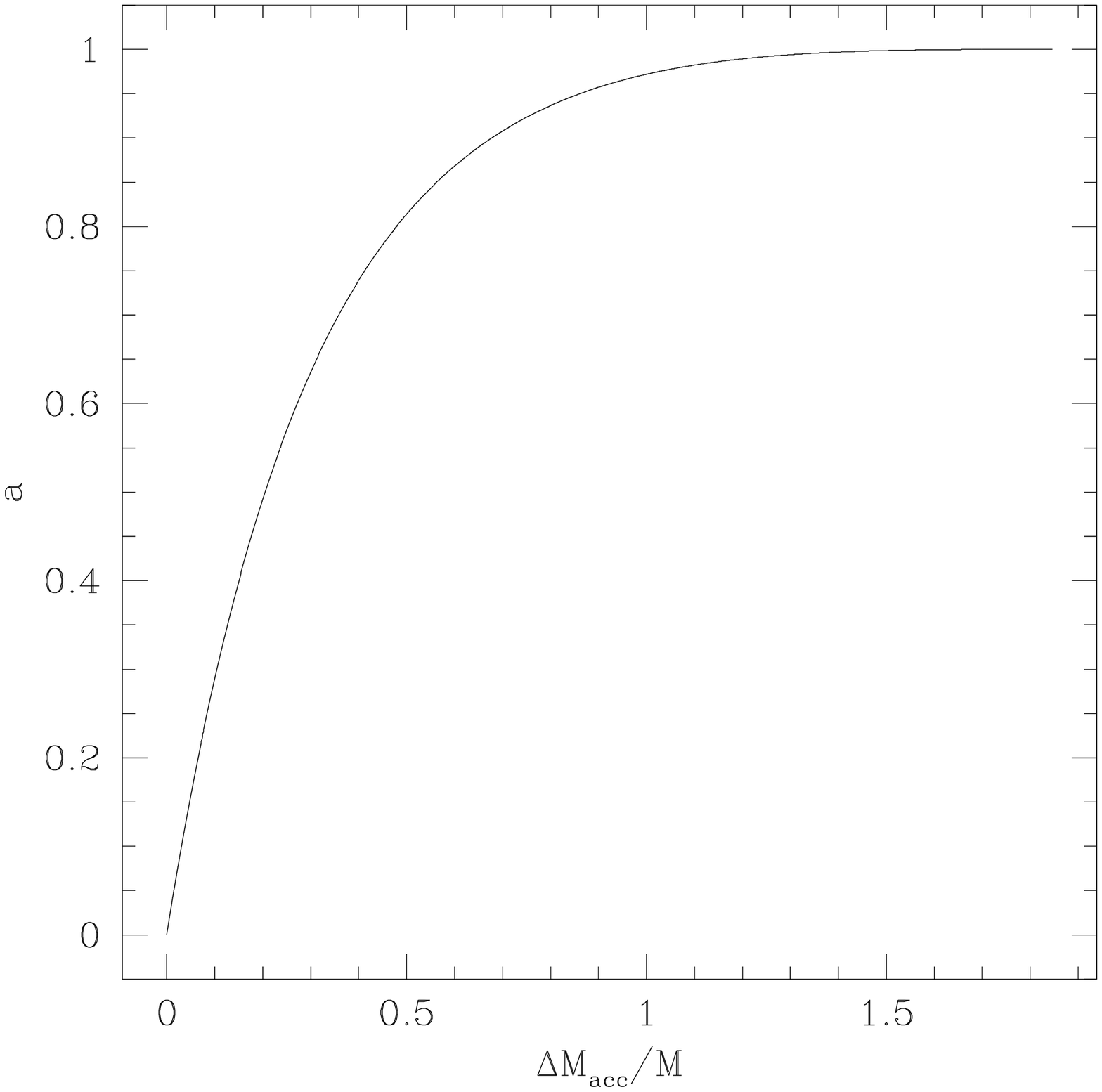}
\epsfysize=8.5truein \epsfbox{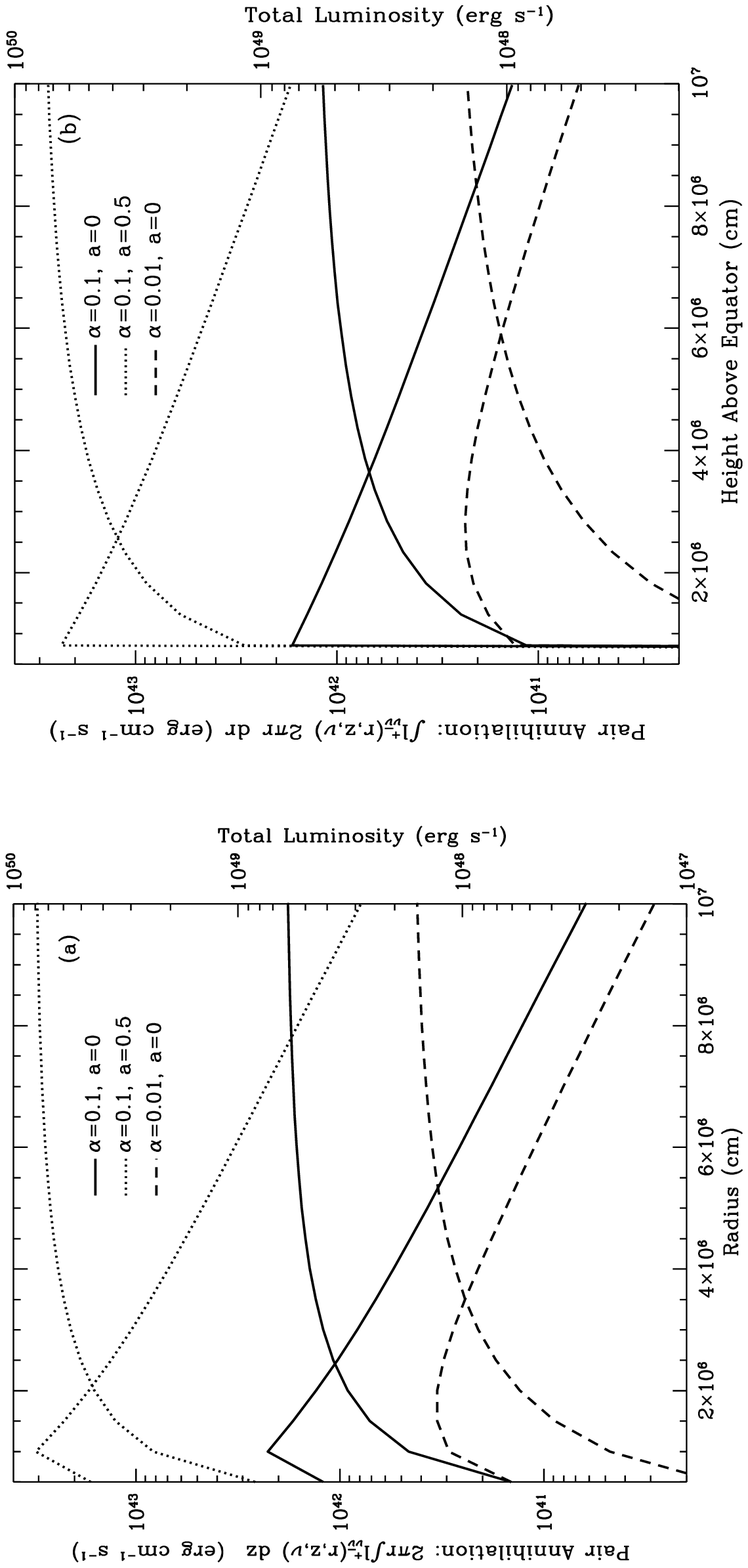}

\end{document}